\documentclass[aps,pra,showpacs,twoside,twocolumn,10pt]{revtex4-1}
\usepackage[colorlinks=true, citecolor=red, urlcolor=blue ]{hyperref}
\usepackage{epsfig,newlfont,amssymb,amsfonts,amsmath,bm,subfigure,palatino,mathtools}
\usepackage{amsthm,braket,times,soul,enumitem,color}
\usepackage[normalem]{ulem}

\begin{document}
 
\title{Estimating  localizable entanglement from witnesses}
\author{David Amaro, Markus M\"uller, and Amit Kumar Pal}
\affiliation{Department of Physics, College of Science, Swansea University, Singleton Park, Swansea - SA2 8PP, United Kingdom}
\begin{abstract}
Computing localizable entanglement for noisy many-particle quantum states is difficult due to the optimization over all possible sets of local projection measurements. Therefore, it is crucial to develop lower bounds, which can provide useful information about the behaviour of localizable entanglement, and which can be determined by measuring a limited number of operators, or by performing least number of measurements on the state, preferably without performing a full state tomography. In this paper, we adopt two different yet related approaches to obtain a witness-based, and a measurement-based lower bounds for localizable entanglement. The former is determined by the minimal amount of entanglement that can be present in a subsystem of the multipartite quantum state, which is consistent with the expectation value of an entanglement witness. Determining this bound does not require any information about the state beyond the expectation value of the witness operator, which renders this approach highly practical in experiments. The latter bound of localizable entanglement is computed by restricting the local projection measurements over the qubits outside the subsystem of interest to a suitably chosen basis. We discuss the behaviour of both lower bounds against local physical noise on the qubits, and discuss their dependence on noise strength and system size. We also analytically determine the measurement-based lower bound in the case of graph states under local uncorrelated Pauli noise.
\end{abstract}
\maketitle

\section{Introduction}
\label{intro}

Over the last two decades, quantum entanglement \cite{horodecki2009} has emerged as a crucial resource in a plethora of quantum information processing tasks, including quantum teleportation \cite{horodecki2009,bennett1993,bouwmeester1997}, quantum dense coding \cite{bennett1992,mattle1996,sende2010}, quantum cryptography \cite{ekert1991,jennewein2000}, and measurement-based quantum computation \cite{raussendorf2001,raussendorf2003,briegel2009}. It has also been proven useful in areas other than quantum information science, such as in detecting quantum phase transitions in quantum many-body systems \cite{osterloh2002,osborne2002,amico2008,chiara2017}, in characterizing topologically ordered states \cite{kitaev2006,pollmann2010,chen2010,jiang2012}, in studying the AdS/CFT correspondence \cite{hubeny2015,pastawski2015,almheiri2015,jahn2017}, and even in areas other than physics, such as in describing the transport properties in photosynthetic complexes \cite{sarovar2010,zhu2012,lambert2013,chanda2014}. Impressive experimental advancement in creating entangled quantum states in the laboratory, by using current technology and substrates such as ions \cite{leibfried2003,leibfried2005,brown2016}, photons \cite{raimond2001,prevedel2009,barz2015}, superconducting qubits \cite{clarke2008,barends2014}, nuclear magnetic resonance molecules \cite{negrevergne2006}, and cold atoms in optical lattices \cite{mandel2003,bloch2005,bloch2008} has enabled the realisation of a wide range of entanglement-based quantum protocols.

Studying the properties of entanglement confined in a subsystem of a increasingly larger multipartite quantum systems remains a pressing task. Many studies aiming at investigating such entanglement have followed two popular approaches. In one, an appropriate entanglement measure is computed for the reduced state $\rho_{N-m}$ of a chosen subsystem $\Omega$ that contains $ N-m $ qubits, obtained by tracing out the $ m $ qubits in the rest of the multipartite system, $\overline{\Omega}$, from the $N$-qubit state $\rho$, such that $\rho_{N-m}=\text{Tr}_{\overline{\Omega}}\rho$ \cite{horodecki2009}. In the other approach, one attempts to obtain entangled post-measurement states over the region $\Omega$ by performing local projection measurements over $\overline{\Omega}$, so that the average entanglement of the states in the post-measurement  ensemble over $\Omega$ is non-zero \cite{divincenzo1998,verstraete2004,popp2005,sadhukhan2017}. For instance, an $N$-qubit Greenberger-Horne-Zeilinger (GHZ) state \cite{greenberger1989} given by $\ket{\mbox{GHZ}}=\frac{1}{\sqrt{2}}\left(\ket{0}^{\otimes N}+\ket{1}^{\otimes N}\right)$ is a classic example where the second approach is particularly useful. Here, the reduced state of $N-m$ qubits for any $m\leq N-2$, given by $\rho_{N-m}^{\text{GHZ}}=\frac{1}{2}\left[(\ket{0}\bra{0})^{\otimes (N-m)}]+(\ket{1}\bra{1})^{\otimes (N-m)}\right]$ has zero entanglement. On the other hand,  the post-measurement states of, say, two qubits, obtained by performing local projection measurements on any one qubit in, say, a three-qubit GHZ state in the basis of Pauli $X$ matrix, are maximally entangled Bell sates $\ket{\phi^{\pm}}=(\ket{00}+\ket{11})/\sqrt{2}$. This motivates one to define \textit{localizable entanglement} as the maximum average entanglement, as measured by an appropriate entanglement measure, localized over $\Omega$ by performing local projection measurements over $\overline{\Omega}$ \cite{verstraete2004,popp2005,sadhukhan2017}. Localizable entanglement has been proven to be indispensable in investigating the correlation length in quantum many-body systems \cite{verstraete2004,popp2005,verstraete2004a,jin2004}, in studying quantum phase transitions in cluster-Ising \cite{skrovseth2009,smacchia2011} and cluster-XY models \cite{montes2012}, in protocols like percolation of entanglement in quantum networks \cite{acin2007}, and in quantifying local entanglement in stabilizer states \cite{raussendorf2003,hein2006,fujii2015,van-den-nest2004}. 

One major challenge with respect to localizable entanglement, even in qubit systems, is its computability, due to the maximization that needs to be performed over all possible local projection measurements on the $m$ measured parties in the $N$-partite system \cite{verstraete2004,popp2005,sadhukhan2017}. Since the number of independent real parameters over which the maximization is to be performed increases with increasing number of measured qubits in the multipartite state \cite{sadhukhan2017}, the computation of localizable entanglement becomes in general difficult even in states of a small number of qubits. Also, in experiments, performing all possible local projection measurements on a set of qubits and determining the post-measurement states by performing state tomography is  resource-intensive and becomes certainly impractical for systems of a large number of qubits. Moreover, an additional complication arises from the fact that one needs to deal with experimental $N$-qubit states which due to noise necessarily deviate to some degree from ideal, often pure target states. In such cases, determination of the localizable entanglement becomes difficult also due to the limited number of computable  measures of entanglement in multipartite mixed states \cite{sadhukhan2017}, if one is interested in localizable entanglement in sets involving more than two qubits. 

In this situation, a promising approach towards understanding the behaviour of localizable entanglement under noise for large stabilizer states is to develop non-trivial as well as computable lower bounds of the actual quantity. This may provide useful information about the system and the dependence of localizable entanglement over different relevant parameters. For example, in the case of the dependence of the localizable entanglement on the noise strength, a non-zero value of the lower bound of the localizable entanglement at a specific value of the noise strength implies sustenance of the actual localizable entanglement for that noise strength. Note that a similar approach of determining computable lower bounds has been adopted in the case of concurrence and entanglement of formation \cite{bennett1996,hill1997,wootters1998,coffman2000,wootters2001}, where the optimization involved in the computation of the actual quantity is difficult to achieve \cite{mintert2004,mintert2004a,mintert2005,mintert2005a,huang2014}. However, in order to satisfy practical purposes, one requires the lower bound of localizable entanglement to be easily computable from limited knowledge of the quantum state, and without performing a full state tomography, for which the required measurement resources increase if the system size is large. It is therefore also imperative  to develop bounds that can be computed by performing least number of local measurements.

There have been attempts to determine the entanglement content and to characterize the dynamics of entanglement in noisy stabilizer states. Methods have been developped in order to obtain lower as well as upper bounds of  entanglement between two subparts in an arbitrarily large graph state  under noise \cite{cavalcanti2009,aolita2010}. Also, the behaviour of long-range entanglement \cite{raussendorf2005}, relative entropy of entanglement \cite{hajdusek2010}, and macroscopic bound entanglement \cite{cavalcanti2010} in cluster states under thermal noise has been investigated. The problem of efficiently estimating relative entropy of entanglement in an experimentally created noisy graph state by stabilizer measurement has also been addressed \cite{wunderlich2010}. Since localizable entanglement is the natural choice for quantifying entanglement between two parties in a multiqubit graph state with or without noise, an in-depth analysis of localizable entanglement in general noisy large-scale graph states is now necessary.

In this paper, we show how computable lower bounds of localizable entanglement can be constructed. For concreteness, we focus on stabilizer states \cite{raussendorf2003,hein2006,fujii2015,van-den-nest2004} and, more specifically, within this class of states, on graph states \cite{hein2006, raussendorf2001,raussendorf2003,briegel2009}, since the characterization of graph states and their properties is well developed and a versatile language for the description of these systems exists. However, since any stabilizer state can be mapped on to a graph state by local unitary operation \cite{van-den-nest2004,hein2006}, our results are either directly translatable, or derivable in a similar way for arbitrary stabilizer states.

We adopt two different, yet related approaches to obtain computable lower bounds for localizable entanglement in the case of mixed quantum states. The first approach is based on entanglement witnesses \cite{terhal2002,guhne2002,bourennane2004,guhne2009,guhne2005,alba2010,amaro} that are local observables whose expectation value signals the presence of entanglement. We use a class of witnesses, called \textit{local} witnesses \cite{guhne2005,alba2010,amaro}, and we show how they can be used to estimate a lower bound of the value of localizable entanglement in subsystems of qubits. Lower bounds of the localizable entanglement can be computed from the expectation values of the witness operators evaluated in the noisy quantum state \cite{brandao2005,brandao2006,eisert2007,guehne2007,guehne2008}. We show that the entanglement measure estimated by the expectation values of these witness operators serve as a faithful lower bound to the actual localizable entanglement on chosen subsystems of specific size. In the second scheme that we explore, we obtain a lower bound of localizable entanglement by considering a specific measurement strategy, thereby restricting the full set of local projection measurement required to compute the localizable entanglement. More specifically, for noisy graph states, we show that a computable lower bound of localizable entanglement is obtained by performing local $Z$ measurements over all qubits in the graph except for the qubits in the region of interest. We establish a relation between these two seemingly unrelated approaches, and test the performance of the obtained lower bounds by benchmarking them for graph states undergoing uncorrelated Pauli noise.

The paper is organized as follows. In Sec. \ref{sec:def}, we  introduce the notation we use and review key concepts of localizable entanglement and graph states, including graph-diagonal states, used throughout this paper. Section \ref{sec:lb} contains a discussion on the local witness-based and local measurement-based lower bounds of localizable entanglement and the interrelation between these bounds. In Sec. \ref{sec:perf}, we demonstrate and compare the performances of the lower bounds in the case of specific noise models, and determine an analytical formula for the measurement-based lower bound in terms of noise-strength and the system size of the analyzed states. Sec. \ref{sec:conclude} contains concluding remarks.


\section{Definitions}
\label{sec:def}

\subsection{Localizable and restricted localizable entanglement}
\label{subsec:le}

The localizable entanglement (LE) \cite{verstraete2004,popp2005,sadhukhan2017} over a number $ N_{\Omega}\geq2 $ of selected qubits forming the region $\Omega$ in a multi-qubit system is defined as the maximum average entanglement that can be accumulated over $\Omega$ by performing local measurements over the qubits in the set $\overline{\Omega}$, where $\Omega\cap\overline{\Omega}=\emptyset$, and $\Omega\cup\overline{\Omega}$ represents the multiqubit system. We denote the state of an $N$-qubit system by $\rho$, where the qubits constituting the system are labelled from $ 1 $ to $ N $ such that $\Omega=\{1,2,3,\cdots,N_{\Omega}\}$, and $\overline{\Omega}=\{N_{\Omega}+1,N_{\Omega}+2,\cdots,N\}$. We label the $m$ ($m=N-N_{\Omega}\leq N-2$) qubits in $\overline{\Omega}$ by $\{r_1,r_2,\cdots,r_m\}$, with $r_i\in\{N_{\Omega}+1,N_{\Omega}+2,\cdots,N\}$, and perform local measurements on them. We restrict ourselves to rank-$1$ projection measurements $\mathcal{M}\equiv\{\mathcal{M}_k;\;k=0,1,2,\cdots,2^{m}-1\}$, in the Hilbert space of $\overline{\Omega}$, which is of dimension $2^{m}$. The post-measurement ensemble $\{p^k,\rho^k_{\Omega}\}$ is represented by the $N$-qubit post-measurement state $\rho^k_{\Omega}$, given by 
\begin{eqnarray}
	\rho^k_{\Omega}=\frac{\mbox{Tr}_{\overline{\Omega}}[\mathcal{M}_k\rho{\mathcal{M}_k}^\dagger]}{\mbox{Tr}[\mathcal{M}_k\rho{\mathcal{M}_k}^\dagger]},
	\label{eq:le-pmstate}
\end{eqnarray}
and the probability with which $\rho^k_{\Omega}$ is obtained, given by 
\begin{eqnarray}
p^k=\mbox{Tr}[\mathcal{M}_k\rho{\mathcal{M}_k}^\dagger].
\end{eqnarray}
Here, $k$ denotes the measurement outcome, and $\sum_{k=0}^{2^{m}-1}p^k=1$. The LE over the $N-m$ qubits in the region $\Omega$ in the $N$-qubit system is given by 
\begin{eqnarray}
E_{\Omega}(\rho)=\underset{\mathcal{M}}{\mbox{sup}}\sum_{k=0}^{2^{m}-1}p^k E(\rho_{\Omega}^k),
\label{eq:localizable_entanglement}
\end{eqnarray} 
where $E$ is a pre-decided entanglement measure. The supremum in Eq.(\ref{eq:localizable_entanglement}) is taken over the complete set of rank-$1$ projection measurements over the qubits in $\overline{\Omega}$.

Rank-$1$ projection measurements on the qubits in $\overline{\Omega}$ can be parametrized as $\mathcal{M}\equiv\{\mathcal{M}_k=\bigotimes_{r_i\in\overline{\Omega}}\ket{k_{r_i}}\bra{k_{r_i}}\}$, where $k_{r_i}\in\{\mathbf{0},\mathbf{1}\}\;\forall r_i\in\overline{\Omega}$, and $\{\ket{k_{r_i}}\}$ are given by \cite{nielsen2010}
\begin{eqnarray}
\ket{\mathbf{0}}_{r_i}&=&\cos (\theta_{r_i}/2)\ket{0}+e^{\text{i}\phi_{r_i}}\sin(\theta_{r_i}/2)\ket{1},\nonumber \\
\ket{\mathbf{1}}_{r_i}&=&\sin (\theta_{r_i}/2)\ket{0}-e^{\text{i}\phi_{r_i}} \cos (\theta_{r_i}/2)\ket{1},
\end{eqnarray}
with $\{\ket{0},\ket{1}\}$ being the computational basis, and $\{(\theta_{r_i},\phi_{r_i}); i=1,2,\cdots,m\}$ are $2m$ real parameters, such that $0 \leq \theta_{r_i} \leq \pi$, $0 \leq \phi_{r_i} < 2\pi$. Here, one can interpret the outcome index $k$ as the multi-index $k_{r_1}k_{r_2}\cdots k_{r_m}$. Therefore, the optimization in Eq.(\ref{eq:localizable_entanglement}) reduces to an optimization over a space of $2m$ real parameters. In general, such optimizations are hard problems when $m$ is large, and can be analytically performed only for a handful of quantum states even in the case of qubit systems \cite{verstraete2004,popp2005,sadhukhan2017}. 

Instead of computing the actual localizable entanglement, one may define a restricted localizable entanglement (RLE) (see \cite{chanda2015} for similar quantities defined in context of quantum information-theoretic measures, such as quantum discord \cite{ollivier2001,henderson2001}), where only single-qubit projection measurements corresponding to the basis of the Pauli operators are allowed. This implies that for each qubit in $\overline{\Omega}$, the possible values of $(\theta_{r_i},\phi_{r_i})$ are \textbf{(i)} $(\theta_{r_i}=0,\phi_{r_i}=0)$ corresponding to the basis $\{\ket{0}_{r_i},\ket{1}_{r_i}\}$ of $Z_{r_i}$, \textbf{(ii)} $(\theta_{r_i}=\pi/2,\phi_{r_i}=0)$ corresponding to the basis $\{\ket{\pm}_{r_i}\}$ of $X_{r_i}$, and \textbf{(iii)} $(\theta_{r_i}=\pi/2,\phi_{r_i}=\pi/2)$ corresponding to the basis $\{\ket{y_{\pm}}_{r_i}\}$ of $Y_{r_i}$, where $\{X,Y,Z\}$ denote the standard Pauli operators.

We denote the complete set of all possible Pauli measurement settings over the $m$ qubits in $\overline{\Omega}$ by $\mathcal{M}^{\mathcal{P}}\equiv\{\mathcal{M}^{\mathcal{P}}_l;l=0,1,2,\cdots,3^m-1\}$. Corresponding to a specific value of $l$, there can be $2^m$ measurement outcomes, denoted by the index $k$, corresponding to each of which the projection operator is given by 
\begin{eqnarray}
\mathcal{M}^{\mathcal{P}}_{(l,k)}=\bigotimes_{r_i\in\overline{\Omega}}\frac{1}{2}\left[I+(-1)^{k_{r_i}}\sigma_{l_{r_i}}\right] 
\label{eq:projform}
\end{eqnarray}
where $l_{r_i}\in\{0,1,2\}$ represents the direction of local projection with $\sigma_{0}=Z$, $\sigma_1=X$, and $\sigma_{2}=Y$ for a specific $r_i$, and $k_{r_i}=0$ $(k_{r_i}=1)$ corresponds to the outcome $+1(-1)$ of the projection measurement. Here, we interpret the index $l$ as the multi-index $l\equiv l_{r_1}l_{r_2}\cdots l_{r_m}$, where the value of $l$ is the base $3$ representation of the string $l_{r_1}l_{r_2}\cdots l_{r_m}$, and the outcome index $k$ as the multi-index $k\equiv k_{r_1}k_{r_2}\cdots k_{r_m}$, where the value of $k$ is the base $2$ representation of the string $k_{r_1}k_{r_2}\cdots k_{r_m}$. Using this notation and following Eq. (\ref{eq:localizable_entanglement}), the RLE is given by 
\begin{eqnarray}
E_{\Omega}^\mathcal{P}(\rho)=\underset{\mathcal{M^{\mathcal{P}}}}{\mbox{sup}}\sum_{k=0}^{2^{m}-1}p^{(l,k)} E(\rho_{\Omega}^{(l,k)}),
\end{eqnarray}
where 
\begin{eqnarray}
\rho_{\Omega}^{(l,k)}&=& \frac{\mbox{Tr}_{\overline{\Omega}}\left[\mathcal{M}^{\mathcal{P}}_{(l,k)}\rho{\mathcal{M}^{\mathcal{P}\dagger}_{(l,k)}}\right]}{\mbox{Tr}[\mathcal{M}^{\mathcal{P}}_{(l,k)}\rho{\mathcal{M}^{\mathcal{P}^\dagger}_{(l,k)}}]},
\label{eq:res_pmstate}
\end{eqnarray}
and 
\begin{eqnarray}
p^{(l,k)}&=&\mbox{Tr}[\mathcal{M}^{\mathcal{P}}_{(l,k)}\rho{\mathcal{M}^{\mathcal{P}\dagger}_{(l,k)}}].
\label{eq:le-pmstate-pauli}
\end{eqnarray}
Clearly, $E_{\Omega}\geq E^\mathcal{P}_{\Omega}$, thereby providing a lower bound to the LE when the optimization is not achieved by Pauli measurements. However, there are important examples and large classes of quantum states, for which $E_{\Omega} = E^\mathcal{P}_{\Omega}$.  These include  \textbf{(i)} graph states \cite{hein2006}, \textbf{(ii)} $N$-qubit generalized GHZ and generalized W states \cite{sadhukhan2017}, \textbf{(iii)} Dicke states and superposition of Dicke states with different excitations and a fixed number of qubits \cite{sadhukhan2017},    \textbf{(iv)} ground states of paradigmatic quantum spin models like the one-dimensional anisotropic $XY$ model in a magnetic field and the $XXZ$ model \cite{verstraete2004,popp2005,venuti2005,sadhukhan2017}, and also \textbf{(v)} the ground states of quantum spin systems described by stabilizer Hamiltonians in the presence of external perturbations in the form of magnetic field or spin-spin interaction, such as the cluster-Ising model \cite{skrovseth2009}.

\subsection{Graph states and stabilizer formalism}
\label{subsec:stab}

A mathematical graph \cite{hein2006,diestel2000,west2001} $\mathcal{G}\left( \mathcal{V},\mathcal{E}\right)$ is composed of a set $ \mathcal{V} $ of $ N $ nodes, labelled by $1,2,\cdots,N-1,N$ and a set $ \mathcal{E} $ of edges or links $(i,j)$ ($i\neq j$) connecting the nodes $i$ and $j$, where $i,j\in\mathcal{V}$. A graph is represented by the adjacency matrix $ \Gamma $, given by 
\begin{eqnarray}
\Gamma_{ij}=\begin{cases}
                         1, & \text{for } (i,j) \in \mathcal{E},  \\
                         0, & \text{for } (i,j) \notin\mathcal{E},
                         \end{cases}
\end{eqnarray} 
which is an $ N\times N $ binary matrix. In this paper, we consider \emph{simple}, \emph{undirected}, and \emph{connected} graphs \cite{hein2006,diestel2000,west2001} only. A simple graph does not contain a loop, i.e., a link connecting a node to itself, and multiple edges between a pair of nodes.  A graph $\mathcal{G}$ is connected if for each pair of sites $ \{i,j\}\in\mathcal{V} $, there exists a path $\mathcal{L}$, constituted of a set of links $\{(k,l)\}\in\mathcal{E}$ with $k,l\in\mathcal{V}$, which connects the nodes $i$ and $j$. Also, in an undirected graph, the links $(i,j)$ and $(j,i)$ are equivalent. We denote the neighbourhood of a node $i$ by $\mathcal{N}_{i}\subset\mathcal{V}$, which is the set of nodes $\{j\}$ in which each node is connected to $i$ by a link, i.e., $(i,j)\in\mathcal{E}$ $\forall$ $j\in\mathcal{N}_i$.  

\begin{figure}
\includegraphics[scale=0.3]{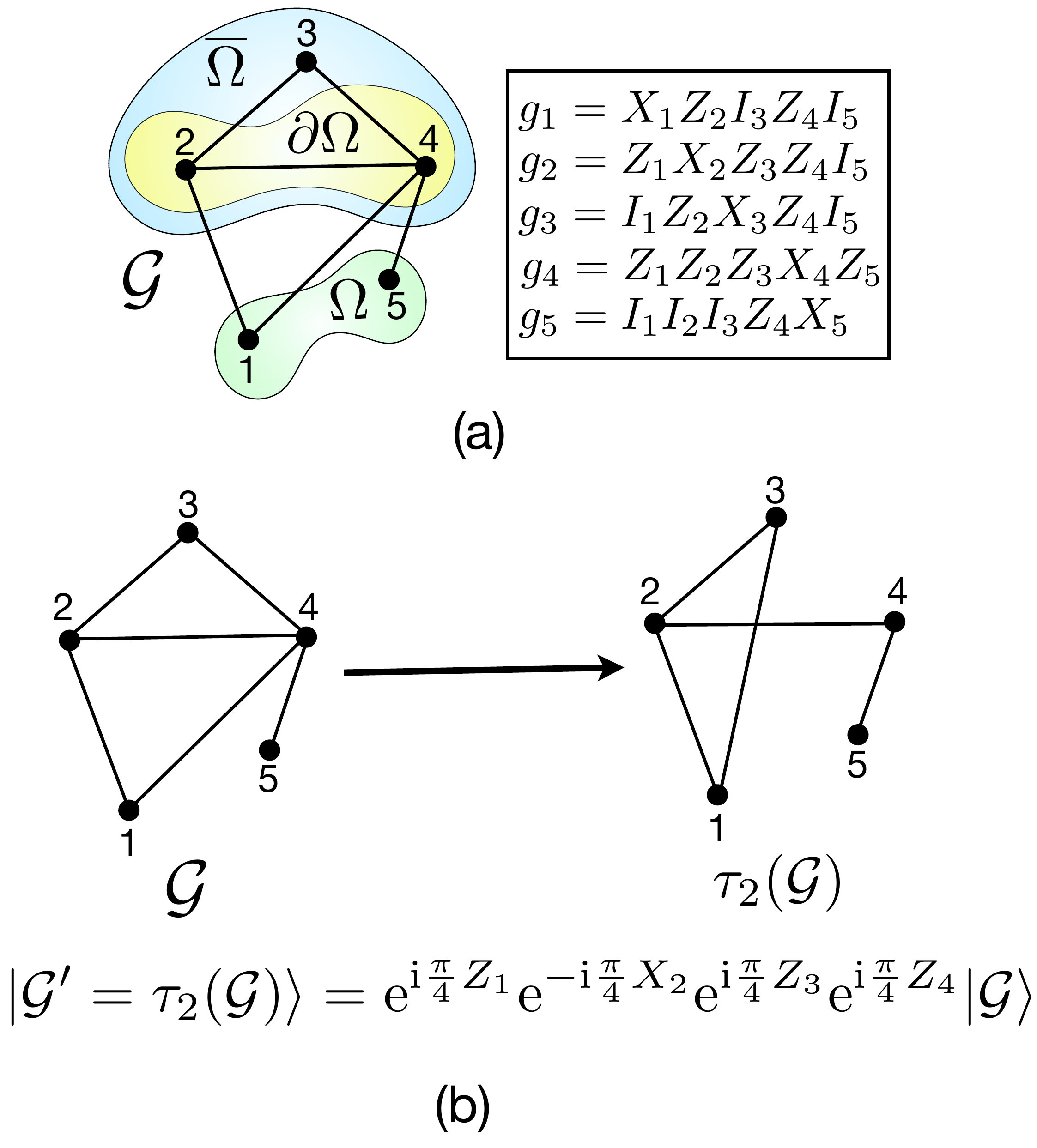}
\caption{{(Color online.) \textbf{Graph state, stabilizers, and local complementation operation.} (a) A five-qubit graph $\mathcal{G}\left( \mathcal{V},\mathcal{E}\right)$, constituted of nodes $\mathcal{V}=\{1,2,3,4,5\}$ and links $\mathcal{E}=\{(1,2),(1,4),(2,3),(2,4),(3,4),(4,5)\}$ is depicted, and the corresponding stabilizer generators $\{g_1,g_2,g_3,g_4,g_5\}$, according to Eq. (\ref{eq:generators}) are explicitly shown. As an example, we consider the subgraph $\mathcal{G}_{\Omega}=(\Omega,\mathcal{E}_{\Omega})$ corresponding to the region $\Omega$ constituted of nodes $\mathcal{V}_{\Omega}=\{1,5\}$ and no links, i.e., $\mathcal{E}_{\Omega}=\emptyset$. On the other hand, $\mathcal{G}_{\overline{\Omega}}=(\overline{\Omega},\mathcal{E}_{\overline{\Omega}})$ is constituted of nodes $\overline{\Omega}=\{2,3,4\}$ and links $\mathcal{E}_{\overline{\Omega}}=\{(2,3),(2,4),(3,4)\}$. The boundary $\partial\Omega$, in this case, is given by $\partial\Omega=\{2,4\}$, and $\mathcal{E}_{\gamma}=\{(1,2),(1,4),(4,5)\}$. (b) A LC operation w.r.t. the node $2$ leads to the graph $\mathcal{G}^\prime$ with modified connectivity, and the corresponding transformation of the graph states, $\ket{\mathcal{G}}\rightarrow\ket{\mathcal{G}^\prime}$ is given by a local unitary transformation according to Eq. (\ref{eq:lcu_def}), as shown explicitly in the figure.}}
\label{fig:graph}
\end{figure}

Let us now consider a region in the graph $\mathcal{G}$, denoted by $\Omega$, which is designated by only the nodes in $\Omega$. For the subgraph $\mathcal{G}_\Omega(\Omega,\mathcal{E}_\Omega)$ corresponding to a region $\Omega$, with $\Omega\subset\mathcal{V}$ and $\mathcal{E}_\Omega\subset\mathcal{E}$, all the above definitions remain valid, and $\mathcal{E}_{\Omega}$ contains only the links $\{(i,j)\}$ such that $i,j\in\Omega$. We denote the cardinality of $\Omega$ by $N_\Omega$ ($ N_\Omega \leq N $). In agreement with the notation used in Sec. \ref{subsec:le}, the rest of the graph is denoted by $\mathcal{G}_{\overline{\Omega}}(\overline{\Omega},\mathcal{E}_{\overline{\Omega}})$, where $\mathcal{E}_{\overline{\Omega}}$ has a definition similar to that of $\mathcal{E}_{\Omega}$ and the set of all nodes is $ \mathcal{V}=\Omega\cup\overline{\Omega} $. The set of links $ \{(i,r_{j})\} $ between a node $ i\in\Omega $ and a node $ r_{j}\in\overline{\Omega} $ is denoted by $ \mathcal{E}_{\gamma} $, so that the complete set of existing links is $ \mathcal{E}=\mathcal{E}_{\Omega}\cup\mathcal{E}_{\overline{\Omega}}\cup\mathcal{E}_{\gamma} $. The boundary $\partial\Omega\subset\overline{\Omega}$ of the region $ \Omega $ is composed by the nodes in $\overline{\Omega}$ that are linked with nodes in $ \Omega $ (see Fig. \ref{fig:graph}(a) for examples of $\mathcal{G}_{\Omega}$, $\mathcal{G}_{\overline{\Omega}}$, $\partial\Omega$, and $\mathcal{E}_{\gamma}$ in a simple graph). Without loss of generality, one can label the nodes such that $\Omega=\{1,2,3,\cdots,N_{\Omega}\}$, and $\overline{\Omega}=\{N_{\Omega}+1,N_{\Omega}+2,\cdots,N\}$, which leads to 
\begin{eqnarray}
\Gamma=\left( \begin{array}{cc} \Gamma_{\Omega} & \gamma^{T} \\
\gamma & \Gamma_{\overline{\Omega}}\end{array}\right). 
\end{eqnarray}
Here, $\Gamma_{\Omega}$ and $\Gamma_{\overline{\Omega}}$ are the adjacency matrices corresponding to $\mathcal{G}_{\Omega}$ and $\mathcal{G}_{\overline{\Omega}}$, respectively, while the $(N-N_{\Omega})\times N_{\Omega}$ matrix $\gamma$ represents the set of links connecting $\Omega$ and $\overline{\Omega}$. In order to keep parity between the notations in Secs. \ref{subsec:le} and \ref{subsec:stab}, we would like to determine the LE over the region $\Omega$ in $\mathcal{G}$, implying $N_{\Omega}=N-m$. 

A graph state $\ket{\mathcal{G}}$ is a multiqubit stabilizer quantum state associated to an undirected graph $\mathcal{G}$, where a qubit is placed at every node in the graph. The state is defined by a set, $G\in\mathcal{P}^{N} $, of mutually commuting generators \cite{hein2006}, $g_i$, where $g_{i}\ket{\mathcal{G}}=\ket{\mathcal{G}}$ $\forall$ $i=1,2,\cdots,N$. Here, $\mathcal{P}^{N}$ denotes the Pauli group \cite{hein2006,nielsen2010}, and the form of the generators $\{g_i\}$, given by
\begin{eqnarray}
g_{i}=X_{i}\otimes\big[\bigotimes_{j\in \mathcal{V}}Z_{j}^{\Gamma_{ij}}\big],
\label{eq:generators}
\end{eqnarray}
is determined by the underlying graph structure (see Fig. \ref{fig:graph}(a) for an explicit example in a five-qubit graph). The generators $\{g_i\}$ share common eigenstates, and the state $\ket{\mathcal{G}}$ is the common eigenstate of $\{g_i\}$ with eigenvalue $+1$. The rest of the $2^N-1$ eigenstates of $\{g_i\}$ are local unitary equivalent to $\ket{\mathcal{G}}$, given by $\{\ket{\mathcal{G}^\nu}=Z_\nu\ket{\mathcal{G}}\}$, where $\nu=0,1,2,\cdots,2^N-1$, and $Z_\nu=\bigotimes_{j\in\mathcal{G}}Z^{\nu_{j}}$, where $\nu_j\in\{0,1\}$. The index $\nu$ is a multi-index $\nu\equiv\nu_1\nu_2\cdots\nu_N$, and can be interpreted as the decimal representation of the binary sequence $\nu_1\nu_2\cdots\nu_N$. In this representation, $\ket{\mathcal{G}}=\ket{\mathcal{G}^0}$. The set of eigenstates $\{\ket{\mathcal{G}^\nu}\}$ forms a complete orthonormal basis of the Hilbert space of the system, and any state that is diagonal in this basis, written as \cite{hein2006,cavalcanti2009,aolita2010,kay2010,kay2011,guhne2011a,guhne2011b}
\begin{eqnarray} 
\rho_{\text{GD}}=\sum_{\nu=0}^{2^N-1}p_\nu\ket{\mathcal{G}^\nu}\bra{\mathcal{G}^\nu},
\label{eq:gdstate}
\end{eqnarray} 
is a graph-diagonal (GD) state, where $\langle \mathcal{G}^\nu |\mathcal{G}^{\nu^\prime}\rangle=\delta_{\nu,\nu^\prime}$, $\delta_{\nu,\nu^\prime}$ being the Kronecker delta, and $\{p_\nu\}$ is any probability distribution. From now on, we shall use the words qubits and nodes interchangeably, and denote them with the same labels, since each node in $\mathcal{G}$ accounts for a specific qubit in $\ket{\mathcal{G}}$.

There exist graph states that are connected to each other by local unitary operations, thereby having identical entanglement properties \cite{hein2006}. A specific set of such states are of particular interest, which correspond to the different graphs connected to each other by the local complementation (LC) operation \cite{hein2006,bouchet1991-93,van-den-nest2004}. The LC operation with respect to a qubit $i$, denoted by $\tau_i(.)$, on a graph $\mathcal{G}$ deletes all the links $\{(j,k)\}$ if $j,k\in\mathcal{N}_i$, and $(j,k)\in\mathcal{E}$, and creates all the links $\{(j,k)\}$ if $j,k\in\mathcal{N}_i$, and $(j,k)\notin\mathcal{E}$.  The operation $\tau_i$ that transforms $\mathcal{G}$ into a new graph $\mathcal{G}'$ is equivalent to a set of local unitary operations, denoted by $U_C^i$, on the corresponding graph state so that $\ket{\mathcal{G}}\rightarrow U_C^i\ket{\mathcal{G}}=\ket{\mathcal{G}'}$, where
\begin{eqnarray}
U_C^i=u_i^x\otimes\left[\bigotimes_{j\in\mathcal{N}_i}u_j^z\right],
\label{eq:lcu_def}
\end{eqnarray}
with $u_i^x=\exp[(-\text{i}\pi/4)X_i]$ and $u_j^z=\exp[(\text{i}\pi/4)Z_j]$ being local Clifford operations (for an example, see Fig. \ref{fig:graph}(b)). For a fixed value of $N$, the set of all possible graphs connected by (sequences of) LC operations over different nodes in the graph is called an orbit \cite{hein2006}. There may exist more than one orbit for a specific value of $N$. The orbits are mutually disjoint sets, and the union of all the orbits corresponding to a fixed value of $N$ provides the complete set of all possible connected graphs.

\section{Lower bounds of localizable entanglement}
\label{sec:lb}

In this section, we establish a relation between the LE over a region $\Omega$ in a graph $\mathcal{G}$ with local entanglement witnesses, and provide a hierarchy of bounds of LE based on suitably chosen local measurements and the expectation values of local entanglement witnesses. 

\subsection{Witness- and measurement-based lower bounds}
\label{subsec:wlb_mlb}

An entanglement witness \cite{terhal2002,guhne2002,bourennane2004,guhne2009,guhne2005,alba2010,amaro} $\mathcal{W}$ is an operator with non-negative expectation values in all separable states, implying that a negative expectation value ($\mathrm{Tr}\left(\rho\mathcal{W}\right)<0$) of the witness operator unambiguously signals the presence of genuine entanglement in $\rho$. A witness operator $\mathcal{W}^g$ that detects the genuine $N$-partite entanglement in a multiparty pure state $\ket{\psi}$ and a state $\rho$ that is close to $\ket{\psi}$ is called a global witness operator, and can be chosen to be of the form \cite{bourennane2004}
\begin{eqnarray}
\mathcal{W}^{g}=\alpha I-\ket{\psi}\bra{\psi}.
\label{eq:global_witness}
\end{eqnarray}   
Here, $I$ is the identity operator in the Hilbert space of $\ket{\psi}$, and $\alpha$ is the largest Schmidt coefficient of $\ket{\psi}$, given by $\alpha=\max_{\{\ket{\phi}\in S_B\}}|\langle\phi |\psi\rangle|^2$, $S_B$ being the complete set of all biseparable states. If $\ket{\psi}$ is a graph state $\ket{\mathcal{G}}$, then it is genuinely multiparty entangled if the underlying graph is connected, and $\mathcal{W}^g$ with $\alpha=\frac{1}{2}$ provides the global entanglement witness operator that can detect entanglement of a noisy state $\rho$ close to the ideal state $\ket{\mathcal{G}}$. Here, $\rho$ may originate from the exposure of an already prepared state $\ket{\mathcal{G}}$ to noise (where we assume that the state $\ket{\mathcal{G}}$ has been prepared with a high fidelity with the actual target state), or in an experiment, where the target state is $\ket{\mathcal{G}}$, but one ends up with a mixed state $\rho$ due to noise in the experimental apparatus. Assuming that the effect of noise in both scenarios can be simulated by known physical noise models,  we consider $\rho=\Lambda(\rho_{\mathcal{G}})$, where $\rho_{\mathcal{G}}=\ket{\mathcal{G}}\bra{\mathcal{G}}$, and the operation $\Lambda(\cdot)$ describes the transformation $\ket{\mathcal{G}}\rightarrow\rho$.

A local witness $\mathcal{W}_{\Omega}$ is an operator that detects the entanglement in a subset $\Omega$ of qubits constituting the state $\rho$. If the subgraph $\mathcal{G}_{\Omega}$ is connected, a local witness can be constructed from the generators $\{g_i\}$ as \cite{guhne2005,alba2010,amaro}
\begin{eqnarray}
\mathcal{W}_{\Omega}=\frac{1}{2} I-\prod_{i\in\Omega}\frac{I+g_{i}}{2},
\label{eq:local_witness}
\end{eqnarray}
with the property that the expectation value of $\mathcal{W}_{\Omega}$ in the state $\rho$ is the same as the expectation value of the witness operator $\mathcal{W}^g_{\Omega}$ in the reduced state $\rho_{\Omega}$, i.e., 
\begin{eqnarray}
\omega=\mathrm{Tr}\left(\rho\mathcal{W}_{\Omega}\right) =\mathrm{Tr}\left( \rho_{\Omega}\mathcal{W}^{g}_{\Omega}\right).  
\label{eq:witness_property}
\end{eqnarray}
Here the witness operator $\mathcal{W}^g_{\Omega}$ is \textit{global} with reference to the region $\Omega$ in $\mathcal{G}$, so that \cite{guhne2005,alba2010,amaro}
\begin{eqnarray}
\mathcal{W}^{g}_{\Omega}=\frac{1}{2}\,I-\ket{\mathcal{G}_{\Omega}}\bra{\mathcal{G}_{\Omega}}, 
\label{eq:witness}
\end{eqnarray}
$\ket{\mathcal{G}_{\Omega}}$ being the graph state corresponding to the subgraph $\mathcal{G}_{\Omega}$. The reduced state $\rho_{\Omega}$ lives only in $\Omega$, and is given by 
\begin{eqnarray}
\rho_{\Omega}=\mathrm{Tr}_{\overline{\Omega}}\left(U_{\gamma} \rho U_{\gamma}^{-1}\right),  
\label{eq:rho_omega}
\end{eqnarray}
where the unitary operator $ U_{\gamma} $ disentangles $ \ket{\mathcal{G}_\Omega} $ from $\ket{\mathcal{G}_{\overline{\Omega}}}$, so that $ U_{\gamma}\ket{\mathcal{G}}=\ket{\mathcal{G}_{\Omega}}\otimes\ket{\mathcal{G}_{\overline{\Omega}}}$ \cite{hein2006}. The unitary operator $U_{\gamma}$, written as
\begin{eqnarray}
U_{\gamma}=\prod_{(i,r_j)\in\mathcal{E}_{\gamma}}U_{ir_j}^{CZ},
\label{eq:u_omega_b}
\end{eqnarray}
is constituted of controlled phase unitaries acting on the links $(i,r_j)\in\mathcal{E}_{\gamma}$ with $i\in\Omega$ and $r_j\in\overline{\Omega}$, given by $ U_{ir_j}^{CZ}=\frac{1}{2}[(I_{r_j}+Z_{r_j})+Z_{i}( I_{r_j}-Z_{r_j})]$. Note here that the operator $\mathcal{W}_{\Omega}$ (Eq. (\ref{eq:local_witness})) is constituted of generators $\{g_i\}$ with $i\in\Omega$. Under the transformation $U_\gamma g_i U_\gamma^{-1}$, the resulting generator no longer has support on $\overline{\Omega}$. Therefore, the unitary operator $U_{\gamma}$  transforms $\mathcal{W}_\Omega$ into $\mathcal{W}_\Omega^g$ as 
\begin{eqnarray}
U_{\gamma}\mathcal{W}_{\Omega}U_{\gamma}^{-1}=\mathcal{W}_{\Omega}^{g}\otimes I_{\overline{\Omega}}. 
\label{eq:witness_unitary}
\end{eqnarray}

Next, we notice that the unitary operator $U_{\gamma}$ is constituted of controlled phase unitaries $U_{ir_j}^{CZ}$ which involve operators $\frac{1}{2}(I_{r_j}\pm Z_{r_j})$ corresponding to the qubits $r_j\in\partial\Omega$ in $Z$. Therefore, writing the identity operator corresponding to the Hilbert space of a specific qubit $r_j\in\overline{\Omega}\mbox{\textbackslash}\partial\Omega$ as $I_{r_j}=[(I_{r_j}+Z_{r_j})+(I_{r_j}-Z_{r_j})]/2$, the form of the unitary operator can be expanded as 
\begin{eqnarray}
U_{\gamma}&=&\sum_{k} \mathcal{Z}^k_{\Omega}\prod_{r_j\in{\overline{\Omega}}}\left(\frac{I+(-1)^{k_{r_j}}Z_{r_j}}{2}\right), 
\label{eq:unitary_expanded}
\end{eqnarray}
where the correction unitaries $\{\mathcal{Z}^k_{\Omega}\}$ are given by
\begin{equation}\label{correction_witnesses}
\mathcal{Z}^{k}_{\Omega}=\prod_{i\in\Omega}Z_{i}^{\textbf{k}\cdot\gamma_{i}},
\end{equation}
where $ \gamma_{i} $ is the $i$-th column of $ \gamma $, $\mathbf{k}$ is a row matrix constituted of the individual measurement outcomes $ k_{r_j} $ corresponding to the qubits $r_{j}\in\overline{\Omega}$, and $\mathbf{u}\cdot\mathbf{v}$ indicates a matrix product calculated modulo $2$ for the matrices $\mathbf{u}$ and $\mathbf{v}$. Note here that $\mathcal{Z}^k_{\Omega}$ acts only on $ \Omega $, and it is determined entirely according to the links in $\mathcal{E}_{\gamma}$, and the values of $\{k_{r_j}\}$ for $r_{j}\in\partial\Omega$. Then, 
\begin{eqnarray}
\rho_{\Omega}&=&\mbox{Tr}_{\overline{\Omega}}(U_{\gamma}\rho  U_{\gamma})
= \sum_{k}p^{(0,k)} \mathcal{Z}_{\Omega}^{k} \rho_{\Omega}^{(0,k)}  \mathcal{Z}_{\Omega}^{k},
\label{eq:relation_mixed}
\end{eqnarray}
where $\rho_{\Omega}^{(0,k)}$ and $p^{(0,k)}$ are for $l=0$ in Eqs. (\ref{eq:res_pmstate}) and (\ref{eq:le-pmstate-pauli}) respectively.

\subsubsection*{Hierarchy of lower bounds}

We are now in a position to establish a hierarchy between a set of quantities that are relevant in investigating the behaviour of localizable entanglement. It is clear from the definition of RLE that although the computational complexity of RLE is less than the same corresponding to a computation of the exact LE, one has in principle still to consider $3^{m}$ possible Pauli measurement settings, which grows exponentially with $m$. For large $m$, where this becomes impractical, one may compute the average entanglement that can be localized on $\Omega$, obtained by choosing a particular setting of Pauli measurement, say, $\mathcal{M}^{\mathcal{P}}_{l}$, in $\overline{\Omega}$, instead of considering the full set of $3^m$ elements of $\mathcal{M}^{\mathcal{P}}$. Here, we have adopted the notation used in Sec.~\ref{subsec:le}. The value of the average entanglement computed in this way depends completely on the choice of the value of $l$. In the scenarios where the choice is not an optimal setting, the average entanglement serves as a lower bound of the RLE, and by extension a lower bound of LE, i.e., 
\begin{eqnarray}
E_{\Omega}(\rho)\geq E^\mathcal{P}_{\Omega}(\rho)\geq E^{l}_{\Omega}(\rho).
\label{eq:ineq}
\end{eqnarray}
We call such a lower bound the \emph{measurement-based lower bound} (MLB) in the following. Unless otherwise stated, throughout this paper, we shall consider Pauli measurements only, and discard the superscript $\mathcal{P}$ from all the operators to keep them uncluttered. Note that a poor choice of the setting may result in vanishing average entanglement corresponding to a trivial lower bound of LE, which highlights the importance of an informed choice of measurement setting from within the full set of Pauli measurements.

In the case of $l=0$, the lower bound $E^0_{\Omega}$ corresponds to local $Z$ measurements on all qubits in $\overline{\Omega}$, and Eq.~(\ref{eq:ineq}) becomes 
\begin{eqnarray} 
E_{\Omega}(\rho)\geq E^\mathcal{P}_{\Omega}(\rho)\geq E^0_{\Omega}(\rho). 
\label{eq:ineq_1}
\end{eqnarray}
A non-zero value of $E^0_{\Omega}$ is likely when $\Omega$ in $\mathcal{G}$ is connected because $\mathcal{M}_0$ is an optimal measurement setting in the absence of noise (i.e., for $\rho=\ket{\mathcal{G}}\bra{\mathcal{G}}$). The use of $E^0_{\Omega}$ as the MLB is justified in scenarios where the state $\rho$ is very close to the graph state $\ket{\mathcal{G}}$, i.e., when the noise acting on the state has very low strength, or when in an experiment the prepared state has very high fidelity with the target state $\ket{\mathcal{G}}$. In such situations, one expects the optimal measurement to not deviate much from the optimal one in the absence of noise. However, in subsequent sections, we shall demonstrate that there exist situations in which $E^0_{\Omega}$ serves as a good choice for MLB even when the noise strength is considerably high.

A clear connection between $E^0_{\Omega}$ and the local entanglement witnesses can now be drawn by using Eq. (\ref{eq:relation_mixed}). The local-unitary invariance of entanglement measures \cite{horodecki2009} implies $E(\mathcal{Z}_{\Omega}^{k}\rho_{\Omega}^{(0,k)}\mathcal{Z}_{\Omega}^{k})=E(\rho_{\Omega}^{(0,k)})$, which leads to  
\begin{eqnarray}
E^0_{\Omega}(\rho)=\sum_{k}p^{(0,k)}E(\mathcal{Z}_{\Omega}^{k}\rho_{\Omega}^{(0,k)} \mathcal{Z}_{\Omega}^{k}), 
\end{eqnarray}
for a specific choice of the entanglement measure $E$. Using the convexity property of entanglement measures \cite{horodecki2009,horodecki2001} results in $E^0_{\Omega}(\rho)\geq E(\rho_{\Omega})$, where $\rho_{\Omega}$ is given by Eq.~(\ref{eq:rho_omega}), and one can modify Eq.~(\ref{eq:ineq_1}) as 
\begin{eqnarray} 
E_{\Omega}(\rho)\geq E^\mathcal{P}_{\Omega}(\rho)\geq E^0_{\Omega}(\rho)\geq E(\rho_{\Omega}). 
\label{eq:ineq_2}
\end{eqnarray}

The quantity $E(\rho_{\Omega})$ may still be difficult to compute in the general case if the region $\Omega$ is large and if $\rho_{\Omega}$ is a mixed state. However, the expectation value $\omega=\mbox{Tr}(\rho_{\Omega}\mathcal{W}_{\Omega}^g)$, which is obtained by measuring $ \mathcal{W}_{\Omega} $ on $ \rho $, can typically be determined, say, in an experiment, with a number of resources that depends only on the size of $ \Omega $, unlike obtaining $\rho_{\Omega}$ from $ \rho $ and the posterior full state tomography for it, which require an effort that depends on the total size of system. From the definition of witness operators, one expects $\omega$ corresponding to a good witness operator and a specific quantum state to be highly negative if the state is highly entangled. Motivated by this, one may use a minimal set of data, and solve an optimization problem which aims to answer the question as to what the minimum amount of entanglement, $E^{\mbox{\scriptsize min\normalsize}}(\rho_{\Omega})$, as measured by any bipartite or multipartite measure $E$, is among all states  $\varrho$, subject to $\varrho$ that are consistent with the data of $\omega$. In other words, one aims to find the quantity given by \cite{eisert2007,guehne2007,guehne2008}
\begin{eqnarray}
E^{\mbox{\scriptsize min\normalsize}}(\rho_{\Omega})=&&\underset{\varrho}{\inf}  E(\varrho),
\end{eqnarray} 
subject to 
\begin{eqnarray}
\omega=\text{Tr}\left(\varrho \mathcal{W}_{\Omega}^g\right)=\mbox{Tr}(\rho_{\Omega}\mathcal{W}_{\Omega}^g),
\end{eqnarray}
where $\varrho$ is in the Hilbert space of $\Omega$, $\varrho\geq 0$, and $\text{Tr}(\varrho)=1$. In the most general scenario, the expectation values of the local witness operators would provide a lower bound of $E^{\mbox{\scriptsize min\normalsize}}(\rho_{\Omega})$, given by $E^{\mathcal{W}}_{\Omega}(\omega)$, so that the inequality in (\ref{eq:ineq_2}) can be further appended as 
\begin{eqnarray} 
E_{\Omega}(\rho)\geq E^\mathcal{P}_{\Omega}(\rho)\geq E^0_{\Omega}(\rho)\geq E(\rho_{\Omega})\geq E^{\mathcal{W}}_{\Omega}(\omega),  
\label{eq:ineq_3}
\end{eqnarray}
where we refer the quantity $E^{\mathcal{W}}_{\Omega}(\omega)$ as the \emph{witness-based lower bound} (WLB) of LE, which is a function of only the expectation value of a local witness $ \omega=\mathrm{Tr}\left( \rho\mathcal{W}_{\Omega} \right) $.

In the following Secs. \ref{subsec:lb_unitary} and \ref{subsec:lb_gdstate} we provide technically detailed discussions of modifications of the hierarchy of lower bounds given in (\ref{eq:ineq_3}) in particular situations, such as under local unitary transformations and for GD states. More specifically, we show that for GD states, $E^0_{\Omega}(\rho)= E(\rho_{\Omega})$, and we use logarithmic negativity \cite{lee2000,vidal2002,plenio2005} as a bipartite entanglement measure to show that for GD states and a region $\Omega$ constituted of two qubits only, $E(\rho_{\Omega})= E^{\mathcal{W}}_{\Omega}(\omega)$. Readers interested in the demonstration of the different lower bounds in the case of graph states under physical noise can skip these discussions, and move on to Sec. \ref{sec:perf}, where we demonstrate the behaviour of the lower bounds under local Pauli noise as functions of the noise strength. 

\subsection{Lower bounds under local unitary transformation}
\label{subsec:lb_unitary}

\begin{figure*}
\includegraphics[width=0.7\textwidth]{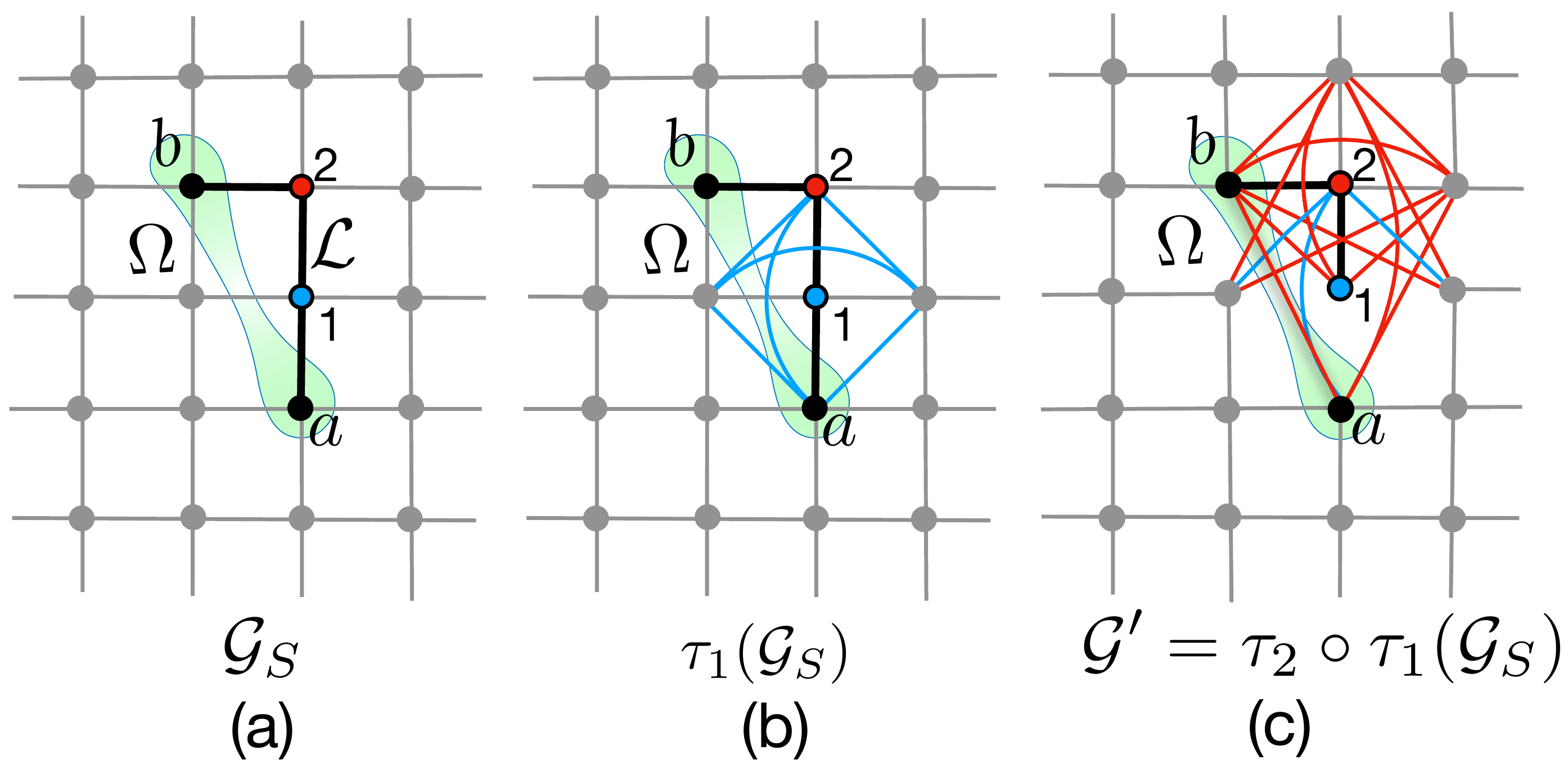}
\caption{(Colour online) \textbf{Creation of a link $(a,b)$ by successive application of LC operations.} (a) A square graph $\mathcal{G}_S$ with a region $\Omega$ of two disconnected qubits $a$ and $b$ denoted by black nodes, joined by a path $\mathcal{L}$, constituted of the qubits $\{a,1,2,b\}$ and the links $\{(a,1),(1,2),(2,b)\}$, denoted by thick black continuous lines. (b) LC operation on qubit ``$1$" (blue)  leading to the graph $\tau_1(\mathcal{G}_S)$). The new links created by the operation are denoted by blue continuous lines. Note that the link $(a,2)$ has been created in this LC operation, which is crucial for the creation of the link $(a,b)$ in the next step. No links are deleted in the operation $\tau_1$.  (c) LC operation on qubit ``$2$" (red) in the graph $\tau_1(\mathcal{G}_S)$ result in the modified graph $\mathcal{G}^\prime=\tau_2\circ\tau_1(\mathcal{G}_S)$, in which the link $(a,b)$ is present. The new links created by this operation are denote by red continuous lines. Note that four of the blue links created in the previous step along with four links from the original graph are deleted by this operation.}
\label{fig:square_graph}
\end{figure*}

An important requirement for the construction of the local witness operator $\mathcal{W}_{\Omega}$ is that the region $\Omega$ in the graph has to be connected. Also, in the case of low noise strength, the value of $E^0_{\Omega}$ can be expected to be non-zero iff $\Omega$ is connected in $\mathcal{G}$, since in the absence of noise, computing $E^0_{\Omega}$ yields zero if $\Omega$ is not connected. However, there may arise situations where the chosen region $\Omega$ in a graph $\mathcal{G}$ is not connected. In that scenario, one may arrive at a graph $\mathcal{G}^\prime$ by performing LC operations over a set of chosen qubits in the graph, so that the region $\Omega$ becomes connected in $\mathcal{G}^\prime$, and the hierarchies given in (\ref{eq:ineq_3}) hold good. For example, let us consider a region $\Omega$ of two disconnected qubits $a$ and $b$. The fact that the original graph $\mathcal{G}$ is connected ensures the existence of a path $\mathcal{L}$ constituted of links $\{(i,j)\}\in\mathcal{E}$ that connects $a$ and $b$. A series of LC operations on the selected qubits $\{i\}\subseteq\mathcal{L}$, where $i\neq a,b$, creates a link between the qubits $a$ and $b$, thereby resulting in a new graph $\mathcal{G}^\prime$ with modified connectivity, where the link $(a,b)$ is present. We illustrate this in Fig. \ref{fig:square_graph} with the example of a square graph. However, a series of LC operations over a graph is equivalent to a local Clifford unitary transformation of the graph state, as demonstrated in Sec. \ref{subsec:stab}. Therefore, in order to check whether Eq.~(\ref{eq:ineq_3}) is valid in the case of a graph where the selected region is not connected, one has to check whether the inequalities remain invariant under such local unitary transformation. 

Remembering that the LC operation on a set of qubits in a graph is equivalent to the application of local Clifford unitaries on a set of qubits in the graph state \cite{van-den-nest2004,hein2006}, without loss of generality, one may write 
\begin{eqnarray} 
\ket{\mathcal{G}^\prime}=U_L\ket{\mathcal{G}}, 
\label{eq:local_unitary}
\end{eqnarray}
where $U_L=\otimes_{i=1}^N U_i$, $\{U_i\}$ being the set of local Clifford unitary operators acting on the qubits $i\in\mathcal{G}$. In the case of a quantum state $\rho$ originating from the graph state due to noise or some error in the experimental setup, without any loss in generality, $\rho^\prime=U_L\rho U_L^{-1}$, where $\rho^\prime$ is the quantum state resulting when $\ket{\mathcal{G}^\prime}$ has undergone the same transformation as $\ket{\mathcal{G}}$ up to the local unitary $ U_{L} $. Note that since $\rho$ and $\rho^\prime$ are connected by local unitary operators, and since LE is invariant under local unitary transformation of the quantum state, $E_{\Omega}(\rho)=E_{\Omega}(\rho^\prime)$ for any connected region $\Omega\in\mathcal{G}$. Moreover, we note that the Clifford unitary operators have the property 
\begin{eqnarray}
\sigma_i =U_i^{-1}\sigma_i^\prime U_i,
\label{eq:clifford_property}
\end{eqnarray}
where both $\sigma_i$ and $\sigma_i^\prime$ are Pauli operators corresponding to the qubit $i$, up to the multiplicative factors $\{\pm 1,\pm\text{i}\}$, while $\sigma_i$ is not necessarily equal to $\sigma^\prime_i$. Since computing the RLE includes all possible Pauli measurement settings, this implies $E_{\Omega}^\mathcal{P}(\rho)=E_{\Omega}^\mathcal{P}(\rho^\prime)$. 

Clearly, the optimal measurement bases for computing LE for $\rho$ and $\rho^\prime$ are not identical. However, the measurement basis corresponding to $\rho$ can be determined by using the knowledge of $U_L$, and an appropriate measurement basis for $\rho^\prime$. In this scenario, we expect $ \rho^\prime $ to be close to the graph state $\ket{\mathcal{G}^\prime}$ where the region $ \Omega $ is connected, so that the appropriate measurement basis for $ \rho^\prime $ should be $\mathcal{M}^0$, which involves only local $Z$ measurement over all qubits in $\overline{\Omega}$. But due to their local unitary connection, the localizable entanglement $ E^0_{\Omega}(\rho^\prime) $ equals $ E^l_{\Omega}(\rho) $, where the value of $l\equiv l_{r_1}l_{r_2}\cdots l_{r_{m}}$ is such that for all $r_i\in\overline{\Omega}$, $\sigma_{l_{r_i}}=U_{L}^{-1}Z_{r_i}U_{L}$, up to the multiplicative factors $\{\pm 1,\pm\text{i}\}$. 

In connection with the local witness operator, one has to now consider
\begin{equation}
\mathcal{W}_{\Omega}^\prime=\frac{1}{2} I-\left( \prod_{i\in\Omega}\frac{I+U_L^{-1} g_{i}^\prime U_L}{2}\right),
\label{eq:wit_unitary}
\end{equation}
with $\{g^\prime_i\}$ being the generators of $\ket{\mathcal{G}^\prime}$ and $ \{U_L^{-1} g_{i}^\prime U_L\} $ are products of the generators $ \{g_{j}\} $ of $\ket{\mathcal{G}}$. Note that the state $\rho_{\Omega}^\prime$ corresponding to $\mathcal{G}^\prime$ is obtained from $ \rho^\prime $ according to Eqs. (\ref{eq:rho_omega}) and (\ref{eq:u_omega_b}), but using a different unitary operator $U_{\gamma^\prime}$, which is defined according to the connectivity of $\mathcal{G}^\prime$. In light of this, the hierarchies of lower bounds in Eq.~(\ref{eq:ineq_3}), in the case of $\mathcal{G}^\prime$, become 
\begin{eqnarray}
E_{\Omega}(\rho)\geq E^\mathcal{P}_{\Omega}(\rho)\geq E^l_{\Omega}(\rho)\geq E(\rho_{\Omega}^\prime)\geq E^{\mathcal{W}}_{\Omega}(\omega^\prime),
\label{eq:ineq_4}
\end{eqnarray}
where $ \omega'=\mathrm{Tr}\left( \rho\mathcal{W}_{\Omega}^\prime\right)  $ and $ \rho_{\Omega}^\prime=\mathrm{Tr}\left( U_{\gamma^\prime}\rho^\prime U_{\gamma^\prime}^{\dagger}\right) $, with $ U_{\gamma^\prime} $ being the disentangling unitary of Eq.~(\ref{eq:u_omega_b}) for $ \ket{\mathcal{G}^\prime} $. 

In scenarios where $\Omega$ is not connected, in the absence of noise, an optimal measurement setting for computing the LE over the region $\Omega$ is the one that corresponds to a sequence of graph operations that results in a connected region $\Omega$. For example, in the case of a disconnected region $\Omega$ constituted of only two qubits, say, ``$a$", and ``$b$", one of the optimal measurement settings corresponds to (i) $X$ measurements on all the qubits that are situated on a path connecting qubits ``$a$" and ``$b$", and (ii) $Z$ measurements on rest of the qubits in the graph \cite{hein2006}. However, there may exist more than one such Pauli measurement setting. Note also that there may exist different sets of local unitary operations that connect $\ket{\mathcal{G}}$ to different graph states where $\Omega$ is connected. Both MLB and WLB described above can therefore be made tighter by considering all such possible cases, and then choosing the maximum of the values.   

\subsection{Lower bounds in graph-diagonal states}
\label{subsec:lb_gdstate}

In this section, we focus on the hierarchies of lower bounds in the case of GD states. The motivation behind determining the structure of lower bounds for GD states stems from the fact that these states occur naturally when graph states are subjected to Pauli noise \cite{cavalcanti2009,aolita2010}, as is demonstrated in Sec. \ref{sec:perf}. Also, any quantum state can be transformed into a GD state by local operations, as demonstrated in \cite{kay2010,kay2011,guhne2011a}. 

Let us first consider the measurement operation $\mathcal{M}_{0}=\{\mathcal{M}_{(0,k)}\}$ with $l=0$ for the $N$-qubit graph state, where the form of $\mathcal{M}_{(l,k)}$ is defined in Eq.~(\ref{eq:projform}) (see Sec.~\ref{subsec:le}). Unless otherwise stated, we keep the value of $l$ fixed at $l=0$ here and throughout the rest of the paper. To keep notation simple, we discard the subscript $l$ from now on, and denote the measurement operation by $\mathcal{M}_0\equiv\{\mathcal{M}_{k}\}$. Here, $\mathcal{M}_{k}=\bigotimes_{r_i\in\overline{\Omega}}\mathcal{M}_{k_{r_i}}$, with $k_{r_i}\in\{0,1\}$. Denoting the graph state as $\rho_{\mathcal{G}}=\ket{\mathcal{G}(\overline{\Omega},\Omega)}\bra{\mathcal{G}(\overline{\Omega},\Omega)}$, implying that $\ket{\mathcal{G}(\overline{\Omega},\Omega)}$ consists of the qubits in $\overline{\Omega}$ and $\Omega$, the effect of operating $\mathcal{M}_{k_{r_i}}$ on $\rho_{\mathcal{G}}$ for a specific $r_i\in\overline{\Omega}$ is given by \cite{hein2006}
\begin{eqnarray}
\hspace{-3mm}\rho_{(\mathcal{G}-r_i)}^{k_{r_{i}}}=\mbox{Tr}_{r_i}\left(\mathcal{M}_{k_{r_i}}\rho \mathcal{M}_{k_{r_i}}\right)
=\frac{1}{2} \mathcal{Z}^{k_{r_i}}\rho_{(\mathcal{G}-r_{i})} \mathcal{Z}^{k_{r_i}}
\label{eq:meas_rule_gen}
\end{eqnarray}
with
\begin{eqnarray}
\mathcal{Z}^{k_{r_i}}=\bigotimes_{j\in\mathcal{N}_{r_i}}Z_j^{k_{r_i}}.
\label{eq:correction}
\end{eqnarray} 
Here, $\mathcal{N}_{r_i}$ represents the neighbourhood of the qubit $r_i$, and $\rho_{(\mathcal{G}-r_i)}=\ket{\mathcal{G}(\overline{\Omega}-r_i,\Omega)}\bra{\mathcal{G}(\overline{\Omega}-r_i,\Omega)}$ corresponds to the graph $\mathcal{G}(\overline{\Omega}-r_i,\Omega)$, obtained from $\mathcal{G}(\overline{\Omega},\Omega)$ by deleting the qubit $r_i$ and all the links that are connected to it. Performing local $Z$-measurement over all qubits in $\overline{\Omega}$, the normalized post-measurement state $\rho_{\mathcal{G}}^{k}$ corresponding to the measurement outcome $k$ can be written as $\rho_{\mathcal{G}}^{k}=\mathcal{Z}^{k} \left( \rho_{\mathcal{G}_{\Omega}}  \otimes \mathcal{M}_k\right) \mathcal{Z}^{k}$, where $\rho_{\mathcal{G}_{\Omega}}=\ket{\mathcal{G}(\Omega)}\bra{\mathcal{G}(\Omega)}$ is the graph state corresponding to the subgraph $\mathcal{G}_{\Omega}$, and the corresponding probability is $p^k=2^{-m}$, which is independent of $k$. The correction is a local operator that can be factorized in a part acting on $ \Omega $ and a part acting on the rest of the qubits, i.e., $ \mathcal{Z}^{k}=\mathcal{Z}^{k}_{\Omega}\otimes\mathcal{Z}^{k}_{\overline{\Omega}} $. Here, $\mathcal{Z}_{\Omega}^{k}$ is the outcome-dependent correction applied to the qubits in $\Omega$ due to the local $Z$ measurements over the qubits in $\overline{\Omega}$ (see Eq.~(\ref{correction_witnesses})). Therefore, tracing out the qubits in $\overline{\Omega}$, the post-measurement state on $\Omega$ corresponding to outcome $k$ is  
\begin{eqnarray} 
\rho_{\mathcal{G}_\Omega}^{k}=\mathcal{Z}_{\Omega}^{k} \rho_{\mathcal{G}_{\Omega}} \mathcal{Z}_{\Omega}^{k}. 
\end{eqnarray}
Similar to Eqs.~(\ref{eq:unitary_expanded}) and (\ref{correction_witnesses}), $ \mathcal{Z}_{\Omega}^{k} $ only depends on the links in $ \mathcal{E}_{\gamma} $.

In the case of GD states, the $N$-qubit post-measurement state, $\rho_{GD}^{k}=\mathcal{M}_k\rho_{GD} \mathcal{M}_k$, corresponding to a specific outcome $k$, can be written as  
\begin{eqnarray}
	{\rho_{GD}^{k}} &=&\sum_{\nu} p_\nu \mathcal{M}_{k} \ket{\mathcal{G}^\nu}\bra{\mathcal{G}^\nu} \mathcal{M}_{k}.
	\label{eq:pm_gdstate}
\end{eqnarray}

Using Eq. (\ref{eq:meas_rule_gen}) in Eq. (\ref{eq:pm_gdstate}), one obtains the normalized post-measurement state corresponding to the outcome $k$ as
\begin{eqnarray}
{\rho_{\text{GD}}^{k}}=\sum_{\nu}p_\nu  Z_{\nu} \mathcal{Z}^{k}\left(  \rho_{\mathcal{G}_{\Omega}}^0\otimes \mathcal{M}_k\right) \mathcal{Z}^{k} Z_\nu, 
\end{eqnarray}
where $\rho_{\mathcal{G}_{\Omega}}^0$ is given by $\rho_{\mathcal{G}_{\Omega}}^0=\rho_{\mathcal{G}_{\Omega}}=\ket{\mathcal{G}(\Omega)}\bra{\mathcal{G}(\Omega)}$. Without loss of generality, we write $Z_{\nu}$ as $Z_{\nu_{\Omega}}\otimes Z_{\nu_{\overline{\Omega}}}$, where the indices $\nu_{\Omega}$ ($\nu_{\Omega}=0,1,2,\cdots,2^{N-m}-1$) and $\nu_{\overline{\Omega}}$ ($\nu_{\overline{\Omega}}=0,1,2,\cdots,2^m-1$) are such that
\begin{eqnarray}
Z_{\nu_{\Omega}}&=&\bigotimes_{i\in\Omega}Z_{i}^{\nu_{\Omega}^i},\nonumber \\
Z_{\nu_{\overline{\Omega}}}&=&\bigotimes_{r_j\in\overline{\Omega}}Z_{r_{j}}^{\nu_{\overline{\Omega}}^{r_j}},
\end{eqnarray}
with $\nu_{\Omega}^i,\nu_{\overline{\Omega}}^{r_j}\in\{0,1\}$. Tracing out the qubits in $\overline{\Omega}$, the post-measurement state ${\rho_{\text{GD},\Omega}^{k}}$ corresponding to the region $\Omega$ can be written as
\begin{eqnarray}
\rho_{\text{GD},\Omega}^{k}=\mathcal{Z}_{\Omega}^{k}\rho_{\text{GD},\Omega}^{0} \mathcal{Z}_{\Omega}^{k}, 
\label{eq:k-dependent}
\end{eqnarray}
with 
\begin{eqnarray}
\rho_{\text{GD},\Omega}^0&=&\sum_{\nu_{\Omega}}\tilde{p}_{\nu_{\Omega}} Z_{\nu_{\Omega}} \rho_{\mathcal{G}_{\Omega}}^0 Z_{\nu_{\Omega}},\nonumber\\
&=&\sum_{\nu_{\Omega}}\tilde{p}_{\nu_{\Omega}}\ket{\mathcal{G}^{\nu_{\Omega}}(\Omega)}\bra{\mathcal{G}^{\nu_{\Omega}}(\Omega)}
\end{eqnarray}
being the post-measurement state corresponding to $k=0$ (i.e., $\mathcal{Z}_{\Omega}^{k}=I_{\Omega}$), where $\tilde{p}_{\nu_{\Omega}^\prime}=\sum_{\nu}p_{\nu}\delta_{\nu_{\Omega},\nu_{\Omega}^\prime}$. Note here that the measurement outcome is reflected only through the correction $\mathcal{Z}_{\Omega}^{k}$. Therefore, the post-measurement states ${\rho_{\text{GD},\Omega}^{k}}$ corresponding to different measurement outcomes $k\neq 0$ are connected to $\rho_{\text{GD}, \Omega}^0$ by local unitary operators of the form $\mathcal{Z}_{\Omega}^{k}$. 
Next, we determine the form of $U_{\gamma}\rho_{\text{GD}}U_{\gamma}^{-1}$, given by 
\begin{eqnarray}
U_{\gamma}\rho_{\text{GD}}U_{\gamma}^{-1}=\sum_{\nu}p_{\nu}Z_\nu U_{\gamma}\rho_{\mathcal{G}}^0 U_{\gamma}^{-1}Z_\nu.  
\end{eqnarray}
Since by the definition of $U_{\gamma}$, $U_{\gamma}\rho_{\mathcal{G}}^0 U_{\gamma}^{-1}=\rho_{\mathcal{G}_{\Omega}}^0\otimes \rho_{\mathcal{G}_{\overline{\Omega}}}^0$, $\rho_{\text{GD},\Omega}=\mbox{Tr}_{\overline{\Omega}}(U_{\gamma}\rho_{\text{GD}}U_{\gamma}^{-1})$ leads to 
\begin{eqnarray}
\rho_{\text{GD},\Omega}=\sum_{\nu_{\Omega}}\tilde{p}_{\nu_{\Omega}} Z_{\nu_{\Omega}} \rho_{\mathcal{G}_{\Omega}}^0 Z_{\nu_{\Omega}}=\rho_{\text{GD},\Omega}^0,
\label{eq:gd_cz}
\end{eqnarray}
with the definitions of $\nu_{\Omega}$ as given above.

We now consider the hierarchy of bounds given in (\ref{eq:ineq_3}), and observe that $E^0_{\Omega}(\rho_{\text{GD}})=E(\rho_{\text{GD},\Omega}^0)$ due to Eq. (\ref{eq:k-dependent}) and the local unitary invariance of entanglement measures. Also, from Eq.~(\ref{eq:gd_cz}), $E(\rho_{\text{GD},\Omega})=E(\rho_{\text{GD},\Omega}^0)$. Combining these observations, the relation in (\ref{eq:ineq_3}) is modified as 
\begin{eqnarray} 
E_{\Omega}(\rho_{\text{GD}})\geq E^\mathcal{P}_{\Omega}(\rho_{\text{GD}})\geq E^0_{\Omega}(\rho_{\text{GD}})= E(\rho_{\text{GD},\Omega})\geq E^{\mathcal{W}}_{\Omega}(\omega_{\text{GD}}) \nonumber \\
\label{eq:ineq_5}
\end{eqnarray}
for GD states, where $ \omega_{\text{GD}}=\mathrm{Tr}\left(\rho_{\text{GD}}\mathcal{W}_{\Omega}\right)   $. 

\subsubsection*{Witness-based lower bound for regions of size two}

We now focus on the WLB in the case of GD states where the region $\Omega$ of interest has size two. For concreteness, we choose logarithmic negativity \cite{lee2000,vidal2002,plenio2005} as the measure of bipartite entanglement. For bipartite quantum states $\varrho_{AB}$ of two parties $A$ and $B$, logarithmic negativity is defined as 
\begin{eqnarray}
L_g(\varrho_{AB})=\log_2(N_g(\varrho_{AB})+1), 
\label{eq:logneg}
\end{eqnarray}
where $N_g(\varrho_{AB})$ is the negativity of $\varrho_{AB}$, based on the Peres-Horodecki separability criterion \cite{peres1996,horodecki1996}, given by  
\begin{eqnarray}
N_{g}=\|\varrho_{AB}^{T_{A}}\|_1-1.
\label{eq:neg}
\end{eqnarray}
Here, $\varrho_{AB}^{T_{A}}$ is the partial transposition of the state $\varrho_{AB}$ with respect to $A$ performed in the computational basis, and $\|\varrho\|_1=\mbox{Tr}\sqrt{\varrho^\dagger \varrho}$ is the trace-norm of $\varrho$. The negativity of the state $\varrho_{AB}$ can then be computed as 
\begin{eqnarray}
N_{g}=2\,\sum_{\lambda_i<0}|\lambda_i|, 
\label{eq:neg_comp}
\end{eqnarray}
where $\{\lambda_i\}$ are the eigenvalues of $\varrho_{AB}^{T_A}$. In the case of witness operators $\mathcal{W}^g_{\Omega}$ given by Eq.~(\ref{eq:witness}), the lower bound $E^{\mathcal{W}}_{\Omega}(\omega)$ of $N_g$, corresponding to a region $\Omega$ of two or three qubits, is given by (see Appendix \ref{ap:wlb_opt})
\begin{eqnarray}
E^{\mathcal{W}}_{\Omega}(\omega) &=&
      \begin{cases}
      -2\omega, & \text{for } \omega < 0,  \\
      0, & \text{for } \omega \geq 0.
      \end{cases}. 
\label{eq:wit_func_form}
\end{eqnarray}
We demonstrate the following results for negativity, which can be straightforwardly extended in the case of logarithmic negativity. 

Using the form of $\rho_{\text{GD},\Omega}$ in Eq.~(\ref{eq:gd_cz}) and the witness operator $\mathcal{W}^g_{\Omega}$ in Eq.~(\ref{eq:witness}), one can determine $\omega_{\text{GD}}=\mbox{Tr}(\rho_{\text{GD},\Omega}\mathcal{W}^g_{\Omega})=\frac{1}{2}-\tilde{p}_0$, implying $E^{\mathcal{W}}_{\Omega}(\omega_{\text{GD}})=2\tilde{p}_0-1$ when $\tilde{p}_0>\frac{1}{2}$ (i.e., $\omega_{\text{GD}}<0$), and $E^{\mathcal{W}}_{\Omega}(\omega_{\text{GD}})=0$ for $\tilde{p}_0\leq\frac{1}{2}$ (i.e., $\omega_{\text{GD}}\geq 0$).

Considering now the two qubits in $\Omega$ to be the two parties $A$ and $B$, $\rho_{\text{GD},\Omega}^{T_{A}}$ is also diagonal in the graph-state basis, similar to $\rho_{\text{GD},\Omega}$, with the eigenvalues of $\rho_{\text{GD},\Omega}^{T_{A}}$ given by 
\begin{eqnarray}
\lambda_{0} &=& 1/2-\tilde{p}_{3},\, \lambda_{1}=1/2-\tilde{p}_{2},\nonumber \\
\lambda_{2} &=& 1/2-\tilde{p}_{1},\, \lambda_{3}=1/2-\tilde{p}_{0}.
\label{eq:eigen_ptran}
\end{eqnarray}
If $\tilde{p}_i\le \frac{1}{2}$ $\forall$ $i\in\{0,1,2,3\}$, $\lambda_i\geq0$, implying $N_g(\rho_{\text{GD},\Omega})=0$. On the other hand, if any of the weights $\{\tilde{p}_i\}$, say $\tilde{p}_j=\max\{\tilde{p}_i\}$ is $>\frac{1}{2}$, then $\tilde{p}_{i\neq j}<\frac{1}{2}$. If $j=0$, then $\lambda_3<0$, implying $N_g(\rho_{\text{GD},\Omega})=2\tilde{p}_0-1$.

Therefore, $N_{g}(\rho_{\text{GD},\Omega})=E^{\mathcal{W}}_{\Omega}(\omega_{\text{GD}})$ if $\tilde{p}_0=\max\{\tilde{p}_{i}\}$, $i=0,1,2,3$, implying that in case of negativity as the entanglement measure, and for $\Omega$ having size two, Eq.~(\ref{eq:ineq_5}) for GD states becomes
\begin{eqnarray} 
E_{\Omega}(\rho_{\text{GD}})\geq E^\mathcal{P}_{\Omega}(\rho_{\text{GD}})\geq E^0_{\Omega}(\rho_{\text{GD}})&=& E(\rho_{\text{GD},\Omega})\nonumber \\ &=& E^{\mathcal{W}}_{\Omega}(\omega_{\text{GD}}).
\label{eq:ineq_6}
\end{eqnarray}
The corresponding logarithmic negativity of $\rho_{\text{GD},\Omega}$ is given by $L_g(\rho_{\text{GD},\Omega})=\log_2(2\tilde{p}_0)$, following Eq.~(\ref{eq:logneg}). In Sec.~\ref{sec:perf}, we consider local, spatially uncorrelated Pauli noise, giving rise to GD states in which $\tilde{p}_0>\frac{1}{2}$ is a common occurrence.

As a final comment, in a region $\Omega$ constituted of two qubits, the bipartite and the genuine multipartite entanglements coincide, but this is not the case if $\Omega$ contains more than two qubits. We shall demonstrate that the use of a bipartite entanglement measure for a region of two qubits results in a tighter WLB where $E_{\Omega}^{\mathcal{W}}$ matches with $E^0_{\Omega}(\rho^\prime)$, while such property is absent when $\Omega$ is bigger (see Fig.~\ref{fig:linear_23}(a)--(b) and the subsequent discussions). The procedure of obtaining a WLB for localizable entanglement over a region $\Omega$ having size bigger than two qubits remains the same as described in Secs. \ref{subsec:wlb_mlb}-\ref{subsec:lb_gdstate} and Appendix \ref{ap:wlb_opt}, the only difference being in the functional form of $E^{\mathcal{W}}_{\Omega}(\omega)$ (Eq. (\ref{eq:wit_func_form})), which depends explicitly on the chosen entanglement measure.  For demonstration, in this paper, we have chosen logarithmic negativity as the measure of bipartite entanglement between the two qubits in $\Omega$ due to the computability of the measure. The main challenge in obtaining a proper WLB for a region $\Omega$ of size larger than two qubits remains in the scarcity of computable genuine multipartite measure of entanglement for mixed multiparty states. However, given such a computable multiparty entanglement measure exists, WLB corresponding to that measure for a region larger than two qubits can be computed by determining $E_{\Omega}^{\mathcal{W}}(\omega)$.

\section{Performance of the lower bounds}
\label{sec:perf}

\begin{figure*}
\includegraphics[width=\textwidth]{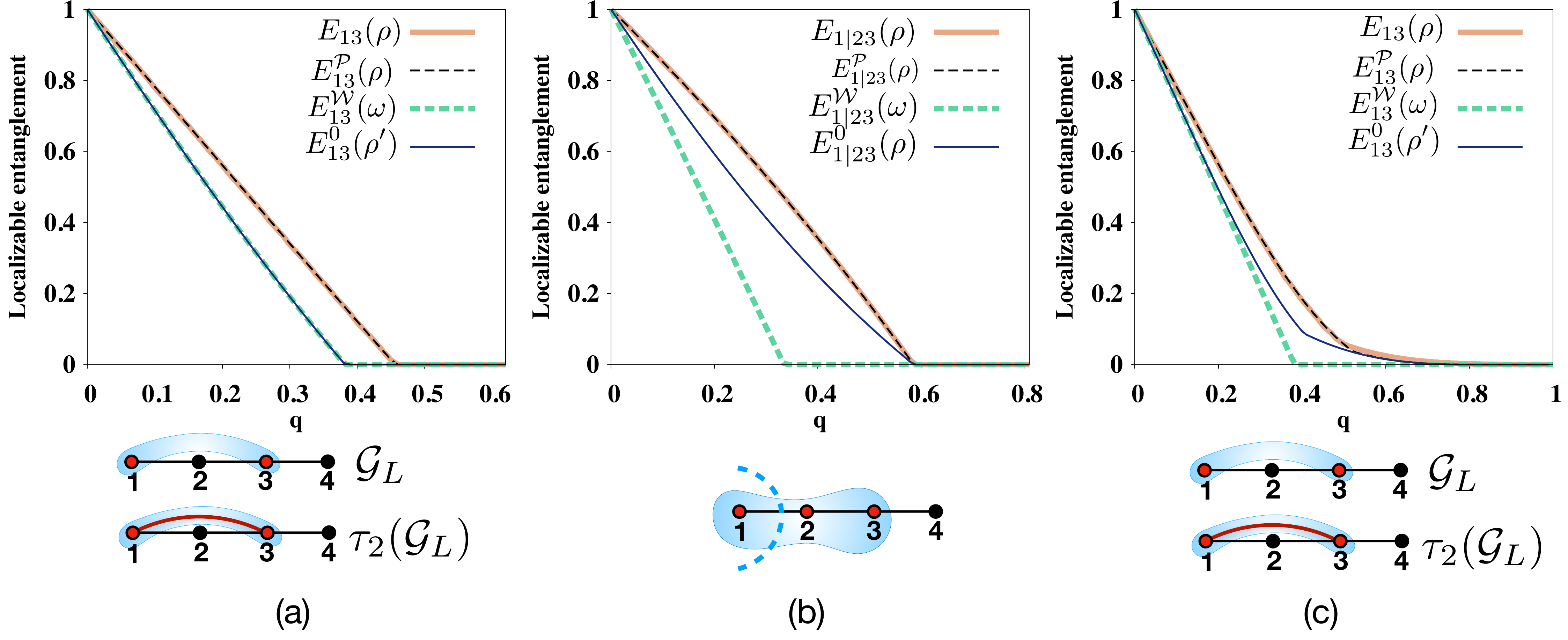}
\caption{(Colour online) \textbf{Localizable entanglement over regions of different size against noise-parameter for linear graphs.} (a) Variations of $E_{13}(\rho)$, $E_{13}^{\mathcal{P}}(\rho)$, $E^0_{13}(\rho^\prime)$ and $E^{\mathcal{W}}_{13}(\omega)$ as functions of the noise parameter $q$ for the region $\Omega\equiv \{1,3\}$ in the linear graph $\mathcal{G}_{L}=\{\mathcal{V}_L,\mathcal{E}_L\}$ composed of four qubits, where $\mathcal{V}_{L}=\{1,2,3,4\}$, and $\mathcal{E}_{L}=\{(1,2),(2,3),(3,4)\}$. We consider BF noise applied to all the qubits. (b) Variations of $E_{1|23}(\rho)$, $E_{1|23}^{\mathcal{P}}(\rho)$, $E^0_{1|23}(\rho)$ and $E^{\mathcal{W}}_{1|23}(\omega)$ as functions of $q$ for the region $\Omega\equiv \{1,2,3\}$ with the bipartition $1|23$ in the linear graph $\mathcal{G}_L$ under BF noise. (c) Variations of $E_{13}(\rho)$, $E_{13}^{\mathcal{P}}(\rho)$, $E^0_{13}(\rho^\prime)$ and $E^{\mathcal{W}}_{13}(\omega)$ as functions of $q$ for the region $\Omega\equiv \{1,3\}$ in the linear graph $\mathcal{G}_{L}$ under AD noise.}
\label{fig:linear_23}
\end{figure*}

In this section, we discuss the performance of the MLB and the WLB discussed in Sec. \ref{sec:lb}. For concreteness, to this end we consider graph states $\mathcal{G}$ under local uncorrelated Pauli noise and local amplitude-damping (AD) noise \cite{nielsen2010}, and discuss how the MLB and the WLB can be computed over a connected region $\Omega$ in the $N$-qubit system. We employ the Kraus operator representation \cite{nielsen2010,cavalcanti2009,aolita2010,holevo2012}, where the evolution of the graph state $\rho_{\mathcal{G}}$ under noise is given by $\rho_\mathcal{G}\rightarrow\rho=\Lambda(\rho_\mathcal{G})$, and where the operation $\Lambda(.)$ can be expressed by an operator-sum decomposition \cite{nielsen2010,holevo2012} given by 
\begin{eqnarray}
\rho = \Lambda(\rho_\mathcal{G}) &=& \sum_{\alpha=0}^{4^N-1}K_\alpha\rho_\mathcal{G} K^\dagger_\alpha \nonumber \\ 
&=& \sum_{\alpha=0}^{4^N-1}q_\alpha J_\alpha\rho_\mathcal{G} J^\dagger_\alpha.
\label{eq:evolve}
\end{eqnarray}
Here, $\{K_\alpha=\sqrt{q_\alpha}J_\alpha\}$ are the Kraus operators satisfying the completeness condition $\sum_{\alpha}K_\alpha^\dagger K_\alpha=I$, with $I$ being the identity operator in the Hilbert space of the system. The map $\Lambda(.)$ in Eq.~(\ref{eq:evolve}) is a completely positive trace-preserving (CPTP) map, and $q$ is the driving parameter of the noise model, which introduces the notion of time, $t$, depending on the type of the physical process through which the system evolves.

For uncorrelated Pauli noise, the individual Kraus operators, $K_{\alpha}$ can be written as the product of identity, $I$, and the three Pauli operators, $X,Y$, and $Z$ acting on the individual qubits. The operators $\{J_{\alpha}\}$ in Eq. (\ref{eq:evolve}) now have the form 
\begin{eqnarray}
J_\alpha=\bigotimes_{i=1}^N\sigma_{\alpha_i},
\label{eq:kraus_pauli}
\end{eqnarray}
and 
\begin{eqnarray}
q_\alpha=\prod_{i=1}^Nq_{\alpha_i},
\label{eq:kraus_prob}
\end{eqnarray}
with $\alpha_i\in\{0,1,2,3\}$, $\sum_{\alpha_i=0}^3q_{\alpha_i}=1$, and $\sigma_{0}=I_i$, $\sigma_{1}=X_i$, $\sigma_{2}=Y_i$, and $\sigma_{3}=Z_i$.  Note here that the index $\alpha$ on the left hand side can be interpreted as the multi-index $\alpha\equiv\alpha_1\alpha_2\cdots\alpha_N$, where $\alpha$ is represented in base $4$ by the string $\alpha_1\alpha_2\cdots\alpha_N$. Examples of Pauli noise include bit-flip (BF), bit-phase-flip (BPF), phase-flip (PF), and depolarizing (DP) channels, with the corresponding values of the probability $q_{\alpha_i}$ given for completeness as follows:
\begin{eqnarray}
\mbox{BF:} &\mbox{\hspace{0.2cm}}& q_{0}=1-\frac{q}{2},q_{1}=\frac{q}{2},q_{2}=0,q_{3}=0; \\
\mbox{BPF:} &\mbox{\hspace{0.2cm}}& q_{0}=1-\frac{q}{2},q_{1}=0,q_{2}=\frac{q}{2},q_{3}=0; \\
\mbox{PF:} &\mbox{\hspace{0.2cm}}& q_{0}=1-\frac{q}{2},q_{1}=0,q_{2}=0,q_{3}=\frac{q}{2}; \\
\mbox{DP:} &\mbox{\hspace{0.2cm}}& q_{0}=1-\frac{3q}{4},q_{1}=\frac{q}{4},q_{2}=\frac{q}{4},q_{3}=\frac{q}{4}.
\label{eq:channel_prob}
\end{eqnarray}
All of these channels induce a complete decoherence on the input quantum state at probability $q = 1$, without any energy exchange with environments, thereby representing non-dissipative noisy channels. Note here that an operation $\sigma_{\alpha_i}$, $\alpha_i=1,2$, on the qubit $i$ of a pure graph state is equivalent to a Pauli $Z$ operator on the qubit $i$ and its neighbourhood, as shown in the following equations:
\begin{eqnarray}
\sigma_{\alpha_i=1} &\leftrightarrow & \bigotimes_{j\in\mathcal{N}_i}Z_j,\nonumber\\
\sigma_{\alpha_i=2} &\leftrightarrow & Z_i\otimes \left[\bigotimes_{j\in\mathcal{N}_i}Z_j\right]. 
\end{eqnarray}
This implies that a graph state under local uncorrelated Pauli noise is a graph-diagonal state \cite{cavalcanti2009,aolita2010}. Hence the discussions in Sec. \ref{subsec:lb_gdstate} apply.  

On the other hand, in the case of local AD noise, the single-qubit Kraus operators are given by 
\begin{eqnarray}
K_0&=&\left(\begin{array}{cc}
   1 & 0 \\
   0 &  \sqrt{1-q}\\
  \end{array}\right),
  K_1=\left(\begin{array}{cc}
   0 & \sqrt{q} \\
   0 &  0\\
  \end{array}\right),
\end{eqnarray}
with $K_2$ and $K_3$ being null operators. Note that although the single-qubit Kraus operators in the case of AD channel can be expanded in terms of Pauli operators, the resulting state $\rho$ due to the application of AD noise to all the qubits in a graph state is not a GD state.

We now illustrate the behaviour of the different quantities in Eq.~(\ref{eq:ineq_3}) for the specific example of a linear graph $\mathcal{G}_{L}=\{\mathcal{V}_L,\mathcal{E}_L\}$ of size $N=4$, where $\mathcal{V}_L=\{1,2,3,4\}$, and $\mathcal{E}_L=\{(1,2),(2,3),(3,4)\}$. We consider two specific cases -- one with a region $\Omega$ of size $2$, constituted of qubits $1$ and $3$ that are not connected by a direct link (see Fig. \ref{fig:linear_23}(a)), and the other with a connected region $\Omega$ of three-qubits, constituted of the qubits $1$, $2$, and $3$. In the first case, one may consider a LC operation on the qubit $2$ to create the link $(1,3)$, so that $\Omega$ becomes connected in the new graph $\mathcal{G}^\prime=\tau_2(\mathcal{G}_L)$. We determine $E_{13}(\rho)$, $E^{\mathcal{P}}_{13}(\rho)$, $E^0_{13}(\rho^\prime)$, and $E^{\mathcal{W}}_{\Omega}(\omega)$ as per the discussions in Sec.~\ref{sec:lb}, when BF noise is applied to all the qubits. Note here that the transformation $\tau_2(.)$ corresponds to the local unitary operation $U_L=\exp (i\pi Z_1/4)\exp(-i\pi X_2/4)\exp(i\pi Z_3/4)$ on $\ket{\mathcal{G}_L}$ (see Sec. \ref{subsec:stab}). Therefore, computing $E^0_{13}(\rho^\prime)$ for the state $\rho^\prime$ is equivalent to computing $E^{l=6}_{13}(\rho)$ for the state $\rho$ by performing $Y$ measurement on the qubit $2$ and $Z$ measurement on the qubit $4$. Recall that the value $l=6$ is the decimal representation of the multi-index $l_{r_1}l_{r_2}$ in base $3$ ($l_{r_1}=2$ for $r_1\equiv 2$, implying $Y$ measurement, and $l_{r_2}=0$ for $r_2\equiv 4$, implying $Z$ measurement), following the notation for measurement bases as introduced in Sec. \ref{subsec:le}. Note also that this differs from the  index convention for designating Pauli operators used in this section. In Fig. \ref{fig:linear_23}(a), we have plotted the variations of $E_{13}(\rho)$, $E^{\mathcal{P}}_{13}(\rho)$, $E^0_{13}(\rho^\prime)$, and $E^\mathcal{W}(\omega)$ as functions of $q$. We observe that irrespective of the structure of the graph, the LE over two and three-qubit regions in graph states under local uncorrelated Pauli noise is always optimized by local Pauli measurements, implying $E_{\Omega}(\rho)=E^{\mathcal{P}}_{\Omega}(\rho)$. Also, in accordance with the results obtained in Sec. \ref{subsec:lb_gdstate}, we find that $E^{0}_{13}(\rho^\prime)=E^{\mathcal{W}}(\omega)$ for all values of $q$. We point out here that the quantity $E^{l=3}_{13}(\rho)$, corresponding to an $X$ measurement on qubit $2$ ($l_{r_1}=1$) and a $Z$ measurement on qubit $4$ ($l_{r_2}=0$), is equal to $E_{13}(\rho)$, as $l=3$ provides the optimal measurement basis in the noiseless case. This is understandable from the fact that the measurement over qubit $2$ commutes with the BF noise applied to it, thereby neutralizing the effect of the noise. This will be discussed in more detail in Sec. \ref{subsec:mlb_arb}.

On the other hand, in the second example, the region of interest $\Omega\equiv\{1,2,3\}$ is already connected. Since we consider a bipartite measure, namely, logarithmic negativity as the measure of entanglement, we focus on the bipartition $1|23$ of the region $\Omega$. However, the results to be reported remain unchanged in the case of other two bipartitions, $2|13$ and $12|3$ also. The variations of $E_{1|23}(\rho)$, $E_{1|23}^{\mathcal{P}}(\rho)$, $E^0_{1|23}(\rho)$, and $E^\mathcal{W}_{1|23}(\omega)$ against the noise parameter $q$ are depicted in Fig. \ref{fig:linear_23}(b). Note here that in contrast to the former example, here $E^0_{1|23}(\rho)>E^\mathcal{W}(\omega)$ for all values of $q$ except at $q=0$, therefore ensuring the validity of the results obtained in Sec.~\ref{subsec:lb_gdstate}. Lastly, we consider the local AD noise as an example of non-Pauli noise, and determine the variations of $E_{13}(\rho)$, $E^{\mathcal{P}}_{13}(\rho)$, $E^0_{13}(\rho^\prime)$, and $E^{\mathcal{W}}_{\Omega}(\omega)$ as functions of $q$. The results are depicted in Fig. \ref{fig:linear_23}(c). The reconstruction of the graph and the corresponding change in the measurement directions are as the same as in Fig. \ref{fig:linear_23}(a).

\subsection{Measurement-based lower bound under Pauli noise for arbitrary graphs}
\label{subsec:mlb_arb}

From the results presented in Fig. \ref{fig:linear_23}(b), it is clear that there exists situations in which $E^{0}_{\Omega}$ may provide a tighter lower bound than $E^\mathcal{W}(\omega)$. However, in the case of noisy graph states of large size, the computation of the quantity $E_{\Omega}^0$ as a lower bound of $E_{\Omega}$ may turn out to be difficult. In this subsection, we shall describe how $E_{\Omega}^0$, in the case of uncorrelated local Pauli noise and a specific connected region $\Omega$, can be computed as a function of the noise parameter, $q$, by using only the knowledge of the connectivity of the underlying graph. For the purpose of demonstration, we consider a region consisting of two qubits $a$ and $b$ only, so that $\Omega\equiv \{a,b\}$. However, the methodology discussed here can be applied to regions of any size in arbitrary graphs.  

Let us consider the general situation where $a$ and $b$ are not connected in $\mathcal{G}$. In such a case, one may obtain a graph $\mathcal{G}^\prime$ with the link $(a,b)$ by the prescriptions discussed in Sec.~\ref{subsec:lb_unitary}. Application of Eq.~(\ref{eq:local_unitary}) in Eq.~(\ref{eq:evolve}) leads to 
\begin{eqnarray}
\rho &=&\sum_{\alpha=0}^{4^N-1}q_\alpha J_\alpha \rho_{\mathcal{G}} J^\dagger_\alpha = U_L^{-1}\rho^\prime U_L,
\label{eq:transform}
\end{eqnarray}  
where
\begin{eqnarray}
\rho^\prime &=& \sum_{\alpha=0}^{4^N-1}q_\alpha J^\prime_{\alpha}  \rho_{\mathcal{G}^\prime} J^\prime_{\alpha}, 
\label{eq:evolve2}
\end{eqnarray}
with $J_{\alpha}^\prime=U_L J_{\alpha}U_L^{-1}$, and $\rho_{\mathcal{G}^\prime}=\ket{\mathcal{G}^\prime}\bra{\mathcal{G}^{\prime}}$. The property of the Clifford operators (Eq.~(\ref{eq:clifford_property})) implies that the operators $J^\prime_{\alpha}=J_{\alpha^\prime}$ in Eq.~(\ref{eq:evolve2}), where $J_{\alpha^\prime}$ is now given by 
\begin{eqnarray}
J_{\alpha^\prime}=\bigotimes_{i=1}^N\sigma_{\alpha_i^\prime}
\end{eqnarray} 
with $\alpha_i^\prime=0,1,2,3$, and $\sigma_{\alpha_i^\prime}=U_i^{-1}\sigma_{\alpha_i}U_i$, where the index $\alpha^\prime$ can be interpreted as the multi-index $\alpha^\prime\equiv\alpha_1^\prime\alpha_2^\prime\cdots\alpha_{N}^\prime$, in the same way as $\alpha$. Note that $\rho^\prime$ is also a GD state.

For reasons that will become clear in the subsequent discussion, we write the modified operators, $\{J_{\alpha^\prime}\}$, and the probabilities $q_\alpha$ in Eq.~(\ref{eq:evolve2}) as
\begin{eqnarray}
J_{\alpha^\prime}&=&J_{\alpha_{ab}^\prime}\otimes\left[\bigotimes_{r_i\in\overline{\Omega}}\sigma_{\alpha_{r_i}^\prime}\right],\nonumber \\
q_\alpha &=& q_{\alpha_{ab}}q_{\overline{\Omega}}
\label{eq:kraus_pauli_2}
\end{eqnarray}
where $J_{\alpha_{ab}^\prime}=\sigma_{\alpha_{a}^\prime}\otimes\sigma_{\alpha_{b}^\prime}$, $q_{\alpha_{ab}}=q_{\alpha_a}q_{\alpha_b}$, and 
\begin{eqnarray} 
q_{\overline{\Omega}}=\prod_{r_i\in\overline{\Omega}}q_{\alpha_{r_i}}.
\end{eqnarray}
Here, $\alpha_{r_i},\alpha_a,\alpha_b \in\{0,1,2,3\}$, and $\sum_{\alpha_i=0}^3q_{\alpha_i}=1$. The index $\alpha^\prime$, $\alpha_{ab}$, and $\alpha_{ab}^\prime$ can be interpreted as the multi-indices $\alpha^\prime\equiv\alpha^\prime_a\alpha^\prime_b\alpha^\prime_{r_1}\cdots\alpha^\prime_{r_{N-2}}$, $\alpha_{ab}\equiv\alpha_a\alpha_b$, and $\alpha_{ab}^\prime\equiv\alpha_a^\prime \alpha_b^\prime$, in the same way as $\alpha$ in Eq.~(\ref{eq:kraus_pauli}).  
Let us now consider the measurement operation $\mathcal{M}_0$, as a result of which the $N$-qubit post-measurement state, ${\rho^\prime}^k=\mathcal{M}_k\rho^\prime \mathcal{M}_k$, corresponding to a specific outcome $k$, can be written as  
\begin{eqnarray}
{\rho^\prime}^k &=&\sum_{\alpha}q_\alpha  J_{\alpha^\prime} \mathcal{M}_{k^\prime} \rho_{\mathcal{G}^\prime} \mathcal{M}_{k^\prime} J_{\alpha^\prime},
\label{eq:pm_state}
\end{eqnarray}
with $\mathcal{M}_{k^\prime}=\bigotimes_{r_i\in\overline{\Omega}}\mathcal{M}_{k^\prime_{r_i}}$, $k^\prime_{r_i}\in\{0,1\}$, where
\begin{eqnarray}
\mathcal{M}_{k_{r_i}^\prime}=\sigma_{\alpha_{r_i}^\prime}\mathcal{M}_{k_{r_i}}\sigma_{\alpha_{r_i}^\prime}.
\label{eq:form_projectors_2_single}
\end{eqnarray} 
The interpretation of the index $k^\prime$ in terms of the indices $\{k_{r_{i}}\}$ corresponding to the outcomes of the measurements on the individual qubits is similar to the other indices, such as $\alpha$, $\alpha^\prime$, $l$, and $k$. Note that the transformation in Eq.~(\ref{eq:form_projectors_2_single}) does not change the basis of the measurement, but changes its outcome.

We proceed along the same line as in Sec.~\ref{subsec:lb_gdstate}, and write the graph state as $\rho_{\mathcal{G}^\prime}=\ket{\mathcal{G}^\prime(\overline{\Omega},a,b)}\bra{\mathcal{G}^\prime(\overline{\Omega},a,b)}$. Use of Eq.~(\ref{eq:meas_rule_gen}) in Eq.~(\ref{eq:pm_state}) over qubits in $\overline{\Omega}$, and then tracing out $\overline{\Omega}$ lead to the two-qubit post-measurement state corresponding to qubits $a$ and $b$, given by 
 \begin{eqnarray}
{\rho^\prime}^k_{ab}&=&\sum_{\alpha}q_{\overline{\Omega}}\left[q_{\alpha_{ab}}J_{\alpha_{ab}^\prime} \rho_{\mathcal{G}_{ab}}^{\beta}J_{\alpha_{ab}^\prime} \right], 
\label{eq:pm_state_ab_2}
\end{eqnarray}
where $\rho_{\mathcal{G}_{ab}}^\beta=\mathcal{Z}^\beta_{ab} \rho_{\mathcal{G}_{ab}} \mathcal{Z}^\beta_{ab}$ and $\rho_{\mathcal{G}_{ab}}=\ket{\mathcal{G}_{ab}}\bra{\mathcal{G}_{ab}}$ is the two-qubit graph state. Here, the set $\{\mathcal{Z}^{\beta}_{ab}=Z^{\beta_a}_{a}\otimes Z_b^{\beta_b}\}$ is constituted of all possible outcome-dependent corrections on $\rho_{\mathcal{G}_{ab}}$ due to different values of $k^\prime$, where $\beta_{a},\beta_{b}\in\{0,1\}$, $Z_{a,b}^0=I_{a,b}$, $Z_{a,b}^1=Z_{a,b}$, and $\beta\equiv \beta_a\beta_b$ is a multi-index given by the decimal representation of the binary string $\beta_a\beta_b$.

Note here that $J_{\alpha_{ab}^\prime}$ and $q_{\alpha_{ab}}$ are independent of the measurement outcome, and depend respectively on the local unitary operator $U_L$ (Eq.~(\ref{eq:evolve2})), and the probability corresponding to the Kraus operators acting on the qubit-pair $(a,b)$ only. Therefore, for a specific graph $\mathcal{G}$, further simplification of the form of the state ${\rho^\prime}^k_{ab}$ is possible by grouping the terms with identical $\rho_{\mathcal{G}_{ab}}^{\beta}$ (i.e., $\rho_{\mathcal{G}_{ab}}^{\beta}$ with the same value of $\beta$) together. Let us introduce the noise local to the qubit pair $(a,b)$ as $\Lambda_{ab}$, where 
\begin{eqnarray}
\Lambda_{ab}(\rho_{\mathcal{G}_{ab}}^\beta)&=&\sum_{\alpha_{ab}}q_{\alpha_{ab}}J_{\alpha_{ab}^\prime}\rho_{\mathcal{G}_{ab}}^\beta J_{\alpha_{ab}^\prime}.
\label{eq:loc_noise}
\end{eqnarray}
Using this notation, Eq.(\ref{eq:pm_state_ab_2}), for a specific graph $\mathcal{G}$, can be written as 
 \begin{eqnarray}
{\rho^\prime}^k_{ab}&=&\sum_{\beta}q_{\overline{\Omega}}^\beta \Lambda_{ab}(\rho_{\mathcal{G}_{ab}}^\beta)=\Lambda_{ab}\big(\tilde{\rho}_{ab}), 
\label{eq:pm_state_ab_3}
\end{eqnarray}
where 
\begin{eqnarray}
\tilde{\rho}_{ab}=\sum_{\beta}q_{\overline{\Omega}}^\beta\rho_{\mathcal{G}_{ab}}^\beta
\label{eq:rhok}
\end{eqnarray}
and for a fixed value of $\beta=\beta^\prime$, $q_{\overline{\Omega}}^{\beta^\prime}$ is the sum of the probabilities $q_{\overline{\Omega}}$ corresponding to all the values of $\alpha$, where $\beta=\beta^\prime$. Note that ${\rho^\prime}^k_{ab}=\tilde{\rho}_{ab}$ iff $\alpha_{a}^\prime=\alpha_b^\prime=0$, implying $\alpha_a=\alpha_b=0$, i.e., qubits $a$ and $b$ are free from noise. If local uncorrelated Pauli noise is present on qubits $a$ and $b$, then the entanglement of the qubit-pair $(a,b)$ decays, implying $E({\rho^\prime}^k_{ab})\leq E(\tilde{\rho}_{ab})$, $E$ being any entanglement measure.  We further note that Eq. (\ref{eq:meas_rule_gen}) suggests that the corrections over the qubit pair $(a,b)$ are fully determined by the neighbourhood of the qubit pair, denoted by $\mathcal{N}_{ab}=\mathcal{N}_a\cup\mathcal{N}_b$, where $\mathcal{N}_{a}(\mathcal{N}_b)$ is the neighbourhood of qubit $a$ ($b$). Therefore, the probability corresponding to the Kraus operator acting on qubit $r_i\notin\mathcal{N}_{ab}$ does not affect the post-measurement state. Since the separability of Pauli maps indicates that $\sum_{\alpha_{{\overline{\Omega}}^\prime}}q_{{\overline{\Omega}}^\prime}=1$ for any ${\overline{\Omega}}^\prime\subset{\overline{\Omega}}$, where $\alpha_{{\overline{\Omega}}^\prime}$ is the multi-index involving the indices $\{\alpha_{r_i}\}$ such that $r_{i}\in{\overline{\Omega}}^\prime$, $q_{\overline{\Omega}}^{\beta}$ can be expressed as 
\begin{eqnarray}
q_{\overline{\Omega}}^{\beta}&=&\sum_{\underset{\overline{\Omega}\in\mathcal{N}_{ab}}{\alpha_{\overline{\Omega}}}}q_{\overline{\Omega}|\beta}.
\label{eq:qr}
\end{eqnarray}

\begin{figure}
\includegraphics[scale=0.4]{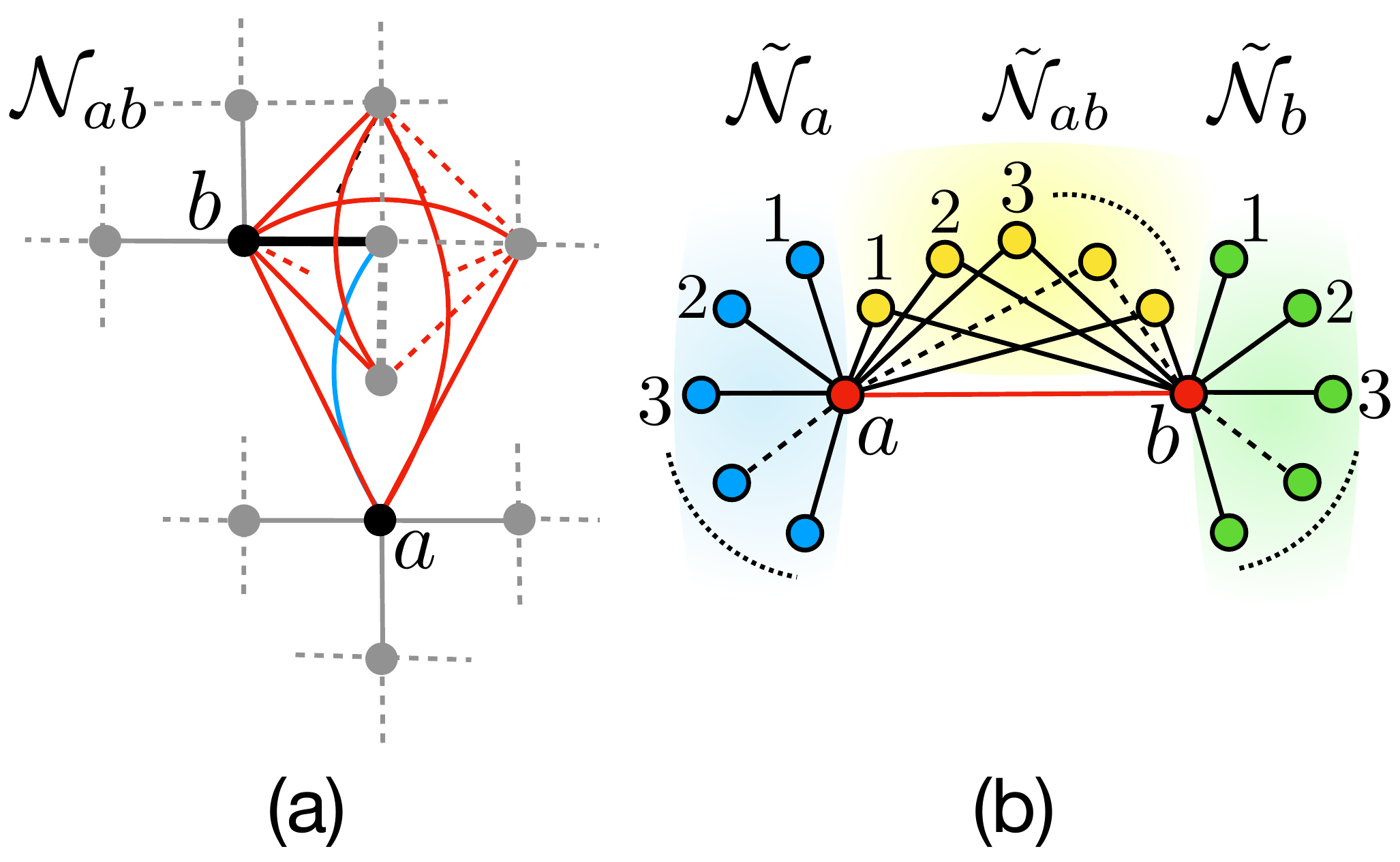}
\caption{(Colour online) \textbf{General structure of the neighbourhood of a connected two-qubit region in an arbitrary graph.} (a) The neighbourhood $\mathcal{N}_{ab}$ of the connected qubits $a$ and $b$ in the graph $\mathcal{G}^\prime=\tau_2\circ\tau_1(\mathcal{G_S})$ shown in Fig.~\ref{fig:square_graph}(b). The links that are connected directly to either of the qubits $a$ or $b$ are depicted by continuous lines, while the links $\{(i,j)\}$ with $i,j\in\mathcal{N}_{ab}$ are represented by broken lines. (b) General structure of $\mathcal{N}_{ab}$ in an arbitrary graph, where the red qubits are the connected qubits of interest, labelled by $a$ and $b$. The neighbourhood  $\mathcal{N}_{ab}$ is constituted of three types of qubits : (1) the qubits that are connected to both $a$ and $b$ (the set $\tilde{\mathcal{N}}_{ab}$, denoted by yellow nodes), (2) the qubits that are connected to only $a$ (the set $\tilde{\mathcal{N}}_a$, denoted by blue nodes), and (3) those connected to only $b$ (the set $\tilde{\mathcal{N}}_b$, denoted by green nodes).}
\label{fig:neigh}
\end{figure}

\subsubsection*{MLB as a function of noise strength and system size}

The dependence of $q_{\overline{\Omega}}^\beta$ on the noise strength and the system size can be explicitly determined by considering a general form of the neighbourhood $\mathcal{N}_{ab}$ in an arbitrary graph $\mathcal{G}^\prime$, where the qubits $a$ and $b$ are connected. Let us consider, for example, the neighbourhood $\mathcal{N}_{ab}$ in the graph $\tau_2\circ\tau_1(\mathcal{G}_S)$ (Fig.\ref{fig:square_graph}(b)). In Fig.~\ref{fig:neigh}(a), we present $\mathcal{N}_{ab}$ corresponding to $\tau_2\circ\tau_1(\mathcal{G}_S)$, where the black (colour online) qubits are the qubits of interest, and $\mathcal{N}_{ab} $ is constituted of the gray (colour online) qubits. The broken links indicate the connectivity of the neighbourhood qubits that are irrelevant in the context of the corrections applied to the qubit pair $(a,b)$ due to local Pauli measurements over the qubits in $\mathcal{N}_{ab}$. On the other hand, the continuous links are the links that connect a qubit in $\mathcal{N}_{ab}$ with either $a$, or $b$, or both, which represent the three types of qubits constituting $\mathcal{N}_{ab}$. Evidently, the corrections on $(a,b)$ according to Eq.~(\ref{eq:meas_rule_gen}) are determined by the connectivity of the qubits in $\mathcal{N}_{ab}$ represented by the continuous links. These features remain unaltered even in the case of a pair of connected qubits in an arbitrary graph.

In Fig. \ref{fig:neigh}(b), we present the most general form of an isolated neighbourhood $\mathcal{N}_{ab}$ of a connected qubit-pair $(a,b)$ in an arbitrary $\mathcal{G}^\prime$. The qubits in $\mathcal{N}_{ab}$ are categorized into three classes according to their connectivity. \textbf{Class 1} consists of the qubits in $\mathcal{N}_{ab}$, denoted by $\tilde{\mathcal{N}}_a$ and represented by the blue (color online) nodes, that are connected to only qubit $a$. The qubits in $\mathcal{N}_{ab}$ that are denoted by $\tilde{\mathcal{N}}_b$ and are connected to only qubit $b$, form the \textbf{Class 2}, and are shown by he green (color online) nodes. And the rest of the qubits in $\mathcal{N}_{ab}$, denoted by $\tilde{\mathcal{N}}_{ab}$, that are connected to both of the qubits $a$ and $b$ is denoted by \textbf{Class 3}. Clearly, $\mathcal{N}_{ab}=\tilde{\mathcal{N}}_a\cup\tilde{\mathcal{N}}_b\cup\tilde{\mathcal{N}}_{ab}$, $\mathcal{N}_{a}=\tilde{\mathcal{N}}_a\cup\tilde{\mathcal{N}}_{ab}$, and $\mathcal{N}_{b}=\tilde{\mathcal{N}}_b\cup\tilde{\mathcal{N}}_{ab}$. From Eq.~(\ref{eq:form_projectors_2_single}), one can also categorize the noise on each qubit in $\mathcal{N}_{ab}$ into two categories. In the first category denoted by \textbf{Type 1}, $k_{r_i}^\prime\neq k_{r_i}$ with a finite probability when the transformation in Eq.~(\ref{eq:form_projectors_2_single}) is carried out (bit-flip and depolarizing channel for example), while $k_{r_i}^\prime$ always equals to $k_{r_i}$ when the noise is of \textbf{Type 2} (for example, phase-flip noise). We denote the set of qubits in $\mathcal{N}_{ab}$ experiencing \textbf{Type 1} (\textbf{Type 2}) noise by $\mathcal{N}_{ab}^1$ ($\mathcal{N}_{ab}^2$), where $\mathcal{N}_{ab}=\mathcal{N}_{ab}^1\cup\mathcal{N}_{ab}^2$, and $\mathcal{N}_{ab}^1\cap\mathcal{N}_{ab}^2=\emptyset$. Similar notations are adopted for qubits in $\tilde{\mathcal{N}}_{a}$, $\tilde{\mathcal{N}}_b$ and $\tilde{\mathcal{N}}_{ab}$ also.  

\begin{figure*}
\includegraphics[width=\textwidth]{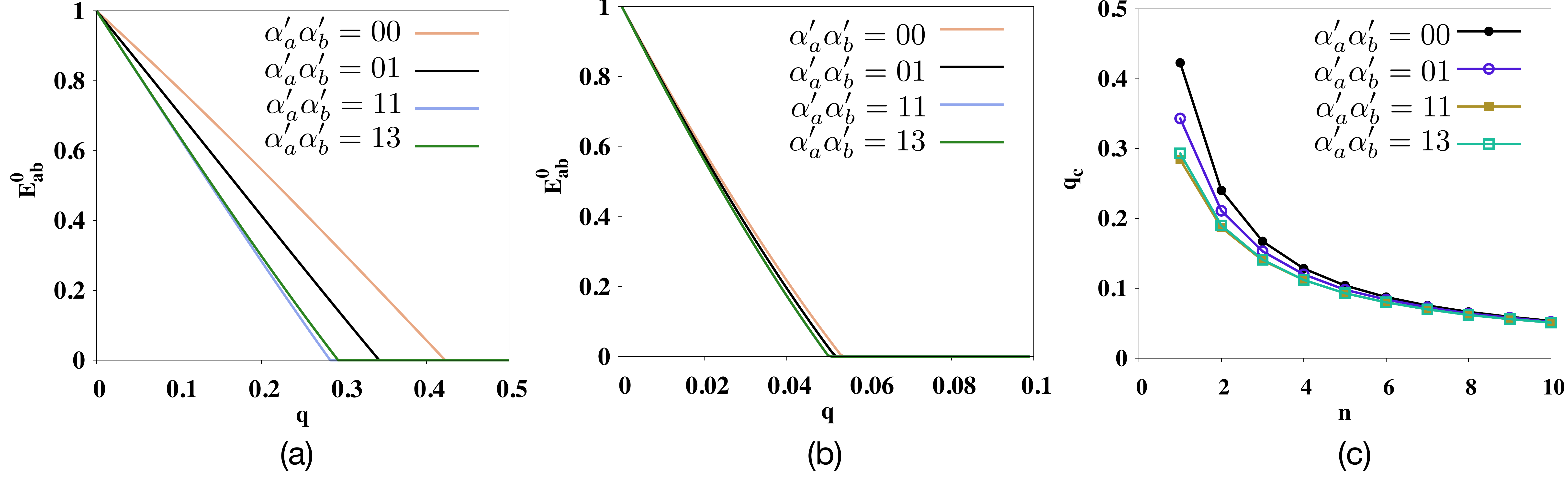}
\caption{(Colour online) \textbf{Measurement-based lower bounds against noise strength for fixed neighbourhood size.} (a) Variation of $E^0_{ab}$ as a function of $q$ for $n_a=n_{ab}=n_b=n$ with $n=1$, when $\alpha_a^\prime\alpha_{b}^\prime=00$ (no noise on qubits $a$ and $b$, Eq.~(\ref{eq:lneg_analytic})), $\alpha_a^\prime\alpha_{b}^\prime=01$ (no noise on qubit $a$ and BF noise on qubit $b$), $\alpha_a^\prime\alpha_{b}^\prime=11$ (BF noise on both qubits $a$ and $b$), and $\alpha_a^\prime\alpha_{b}^\prime=13$ (BF noise on qubit $a$ and PF noise on qubit $b$). (b) Variation of $E^0_{ab}$ as a function of $q$ for $n_a=n_{ab}=n_b=n$ with $n=10$, when $\alpha_a^\prime\alpha_{b}^\prime=00$, $\alpha_a^\prime\alpha_{b}^\prime=01$, $\alpha_a^\prime\alpha_{b}^\prime=11$, and $\alpha_a^\prime\alpha_{b}^\prime=13$. (c) Variation of $q_c$ as a function of $n$ in the case of $\alpha_a^\prime\alpha_{b}^\prime=00$ (Eq.~(\ref{eq:pcrit})), $01$, $11$, and $13$.}
\label{fig:qc}
\end{figure*}

Let us first determine the form of ${\rho^\prime}^k_{ab}$ when only the set $\mathcal{N}_{ab}^1$ is populated, and $\mathcal{N}_{ab}^2=\emptyset$. Non-zero contribution in $q_{\overline{\Omega}}^\beta$ is provided by the qubits in $\mathcal{N}_{ab}^1$ due to the probabilistic change of the outcome from $k_{r_i}$ to $k_{r_i}^\prime$, along with the application of appropriate corrections $\mathcal{Z}^{\beta}_{ab}$ on $\rho_{\mathcal{G}_{ab}}$. Without loss of generality, let us denote the number of qubits in $\tilde{\mathcal{N}}_{a}^1$, $\tilde{\mathcal{N}}_{b}^1$, and $\tilde{\mathcal{N}}_{ab}^1$ by $n_a$, $n_b$, and $n_{ab}$, respectively. 
Let us also assume that corresponding to a specific outcome $k$ in Eq.~(\ref{eq:pm_state}), $n^{0}_a$ of the outcomes $\{k_{r_i};r_i\in\overline{\Omega}\in\tilde{\mathcal{N}}_{a}^1\}$ are $0$, while $n_a^1$ are $1$, such that $n_a=n_a^0+n_a^1$. Similar definitions apply for $n_{b}^{0,1}$ and $n_{ab}^{0,1}$. Interpreting $q_{\overline{\Omega}}^\beta$ as the probability that the correction $\mathcal{Z}^{\beta}_{ab}$ is applied to $\rho_{\mathcal{G}_{ab}}$, its explicit form can be determined as (see Appendix \ref{ap:pmstate} for a detailed derivation)
\begin{eqnarray}
q_{\overline{\Omega}}^{0} &=& P_{a}^- P_{ab}^-P_{b}^- + P_{a}^+ P_{ab}^+ P_{b}^+,\nonumber\\
q_{\overline{\Omega}}^{1} &=& P_{a}^- P_{ab}^-P_{b}^+ + P_{a}^+ P_{ab}^+ P_{b}^-,\nonumber\\
q_{\overline{\Omega}}^{2} &=& P_{a}^+ P_{ab}^-P_{b}^- + P_{a}^- P_{ab}^+ P_{b}^+,\nonumber\\
q_{\overline{\Omega}}^{3} &=& P_{a}^- P_{ab}^+P_{b}^- + P_{a}^+ P_{ab}^- P_{b}^+,
\label{eq:mix_prob}
\end{eqnarray}
with
\begin{eqnarray}
P_a^\pm &=&\frac{1}{2}\left[1\pm (-1)^{n_{a}^1}(1-q)^{n_a}\right],\nonumber\\
P_b^\pm &=&\frac{1}{2}\left[1\pm (-1)^{n_{b}^1}(1-q)^{n_b}\right],\nonumber \\ 
P_{ab}^\pm &=&\frac{1}{2}\left[1\pm (-1)^{n_{ab}^1}(1-q)^{n_{ab}}\right],
\label{eq:expl}
\end{eqnarray}
where we have assumed the noise to be of BF, BPF, or DP type. Therefore, ${\rho^\prime}^k_{ab}$ (Eq.~(\ref{eq:pm_state_ab_3})), in its explicit form, can be determined as a function of the size of $\mathcal{N}_{ab}^1$ and $q$ by using Eqs.~(\ref{eq:mix_prob})-(\ref{eq:expl}) as ${\rho^\prime}^k_{ab}=\Lambda_{ab}(\tilde{\rho}_{ab})$ with 
\begin{eqnarray}
\tilde{\rho}_{ab}&=&q_{\overline{\Omega}}^0\mathcal{Z}^0_{ab}\rho_{\mathcal{G}_{ab}}\mathcal{Z}^0_{ab}+q_{\overline{\Omega}}^1\mathcal{Z}^1_{ab}\rho_{\mathcal{G}_{ab}}\mathcal{Z}^1_{ab}\nonumber \\
&&+q_{\overline{\Omega}}^2\mathcal{Z}^2_{ab}\rho_{\mathcal{G}_{ab}}\mathcal{Z}^2_{ab}+q_{\overline{\Omega}}^3\mathcal{Z}^3_{ab}\rho_{\mathcal{G}_{ab}}\mathcal{Z}^3_{ab},
\label{eq:final_form}
\end{eqnarray}
where the form of $\Lambda_{ab}$ is given in Eq.~(\ref{eq:loc_noise}). In the general scenario where $\mathcal{N}_{ab}^2\neq\emptyset$, its only contribution to ${\rho^\prime}_{ab}^k$ is an extra correction belonging to the set $\{\mathcal{Z}^\beta_{ab}\}$ according to the connectivity of the qubits in $\mathcal{N}_{ab}$. However, $\mathcal{Z}^\beta_{ab}$ being a local unitary operator, the entanglement properties of ${\rho^{\prime}}^k_{ab}$ remain unchanged, and Eq.~(\ref{eq:expl}) represents the effective form of ${\rho^\prime}^k_{ab}$ as far as entanglement is concerned. Therefore, the dependence of the entanglement of ${\rho^{\prime}}^k_{ab}$ on the noise strength and the size of the system is solely determined by the qubits in $\mathcal{N}_{ab}^1$. Note here that the two-qubit post-measurement states corresponding to different values of $k$ are connected by local unitary operators (see Sec.~\ref{subsec:lb_gdstate}), implying that it is sufficient to consider ${\rho^\prime}^0_{ab}$, or any other value of $k$, since $E_{ab}^0(\rho^\prime)=E({\rho^\prime}^k_{ab})=E({\rho^\prime}^0_{ab})$ (see Eq.~(\ref{eq:ineq_5})).

To investigate the features of the MBL as a function of the noise strength and the system size, we choose logarithmic negativity as the measure of bipartite entanglement, $E$. From the expression of ${\rho^\prime}_{ab}^{k}$ (Eq. (\ref{eq:final_form})), it is clear that $L_g({\rho^\prime}_{ab}^{k})\leq L_g(\tilde{\rho}_{ab})$ (see Eqs. (\ref{eq:pm_state_ab_3}) -- (\ref{eq:rhok}) and subsequent discussions). For the purpose of demonstration, we consider the scenario where noise is absent on qubits $a$ and $b$, i.e., ${\rho^\prime}^k_{ab}=\tilde{\rho}_{ab}$.
One can compute the logarithmic negativity of the state $\tilde{\rho}_{ab}$ from Eq.~(\ref{eq:logneg}). The negativity of the state $\tilde{\rho}_{ab}$, for a fixed value of $q$ is given by Eq. (\ref{eq:neg_comp}), where $\{\lambda_i;i=0,1,2,3\}$  are the eigenvalues of $\tilde{\rho}_{ab}^{T_a}$. These eigenvalues can be explicitly computed in a similar fashion as in Eq.~(\ref{eq:eigen_ptran}) by identifying $\tilde{p}_i$ to be equivalent to $q^\beta_{\overline{\Omega}}$, where both $i,\beta=0,1,2,3$. As functions of $q$, $n_a$, $n_b$, and $n_{ab}$, $\{\lambda_i\}$ are given by
\begin{eqnarray}
\lambda_0 &=&\frac{1}{4}[1+\tilde{q}^{n_a+n_{ab}}-\tilde{q}^{n_a+n_b}+\tilde{q}^{n_{ab}+n_{b}}],\nonumber\\
\lambda_1 &=&\frac{1}{4}[1+\tilde{q}^{n_a+n_{ab}}+\tilde{q}^{n_a+n_b}-\tilde{q}^{n_{ab}+n_{b}}],\nonumber\\
\lambda_2 &=&\frac{1}{4}[1-\tilde{q}^{n_a+n_{ab}}+\tilde{q}^{n_a+n_b}+\tilde{q}^{n_{ab}+n_{b}}],\nonumber\\
\lambda_3 &=&\frac{1}{4}[1-\tilde{q}^{n_a+n_{ab}}-\tilde{q}^{n_a+n_b}-\tilde{q}^{n_{ab}+n_{b}}],
\end{eqnarray}
where $q+\tilde{q}=1$. For the purpose of illustration, let us now consider the situation where $n_{a}=n_{ab}=n_b=n$. In this case, the eigenvalues of $\tilde{\rho}_{ab}^{T_a}$ are $\lambda_0=\lambda_1=\lambda_2=\frac{1}{4}[1+\tilde{q}^{2n}]$, and $\lambda_{3}=\frac{1}{4}[1-3\tilde{q}^{2n}]$, of which the negative eigenvalue is $\lambda_{3}$ in the range $0\leq q <1-\left(\frac{1}{3}\right)^{\frac{1}{2n}}$. In this range, $E^0_{ab}$ as a function of $q$ and $n$ can be expressed as 
\begin{eqnarray}
E^0_{ab}=\log_{2}\left[3(1-q)^{2n}+1\right]-1. 
\label{eq:lneg_analytic}
\end{eqnarray}
For a specific value of $n$, $E^0_{ab}$ goes to zero at a critical value 
\begin{eqnarray}
q_c=1-\left(\frac{1}{3}\right)^{\frac{1}{2n}}.
\label{eq:pcrit}
\end{eqnarray}
For $q>q_c$, $\lambda_3$ becomes positive, and the logarithmic negativity vanishes.

In Fig.~\ref{fig:qc}(a), we plot the variation of $E^0_{ab}$ as a function of the noise strength $q$ with $n=1$, for different types of noise present on the qubit pair $(a,b)$. We conveniently denote the different types of noise on $(a,b)$ by the multi-index $\alpha_{ab}^\prime\equiv\alpha_a^\prime\alpha_{b}^\prime$, where, for example, $\alpha_a^\prime\alpha_b^\prime=11$ implies bit-flip noise applied to both qubits $a$ and $b$. We find that the variation of $E^0_{ab}$ with $q$ in the case of $\{\alpha_a^\prime\alpha_b^\prime=01,02,03,10,20,30\}$ are quantitatively identical. Similar behaviour is observed in the case of $\{\alpha_a^\prime\alpha_b^\prime=11,12,21,23,32,33\}$ and $\{\alpha_a^\prime\alpha_b^\prime=13,22,31\}$. With an increase in the value of $n$, the value of $E_{ab}^0$ for a fixed value of $q$ decreases, and the effect of the noise on the region $\Omega\equiv \{a,b\}$ becomes less prominent. This is clearly shown by the coincidence of the variations of $E^0_{ab}$ against $q$, when the neighbourhood size is increased to $n=10$ (see Fig.~\ref{fig:qc}(b)). The variation of $E^0_{ab}$ with $q$ remains qualitatively unchanged if one considers different relations between $n_a$, $n_{ab}$, and $n_b$ instead of $n_a=n_{ab}=n_b=n$. However, identical dynamics is now shown by groups of noise channels, denoted by specific values of $\alpha_a^\prime\alpha_b^\prime$,  which are different from that in the former case.  In Fig.~\ref{fig:qc}(b), we plot the variation of $q_c$ as a function of increasing $n$ for different types of noise on the qubits $a$ and $b$, where the data for $\alpha_a^\prime\alpha_{b}^\prime=00$ corresponds to Eq.~(\ref{eq:pcrit}), and the data corresponding to the rest of the noise models are obtained numerically, by considering $E^0_{ab}=0$ for values below a numerical cut-off, concretely, if $E_{ab}^0<10^{-6}$. The qualitative behaviour of $q_c$ against the system size is found to remain invariant for different relations between $n_a$, $n_{ab}$, and $n_b$ instead of $n_a=n_{ab}=n_b=n$.

\begin{figure}
\includegraphics[scale=0.4]{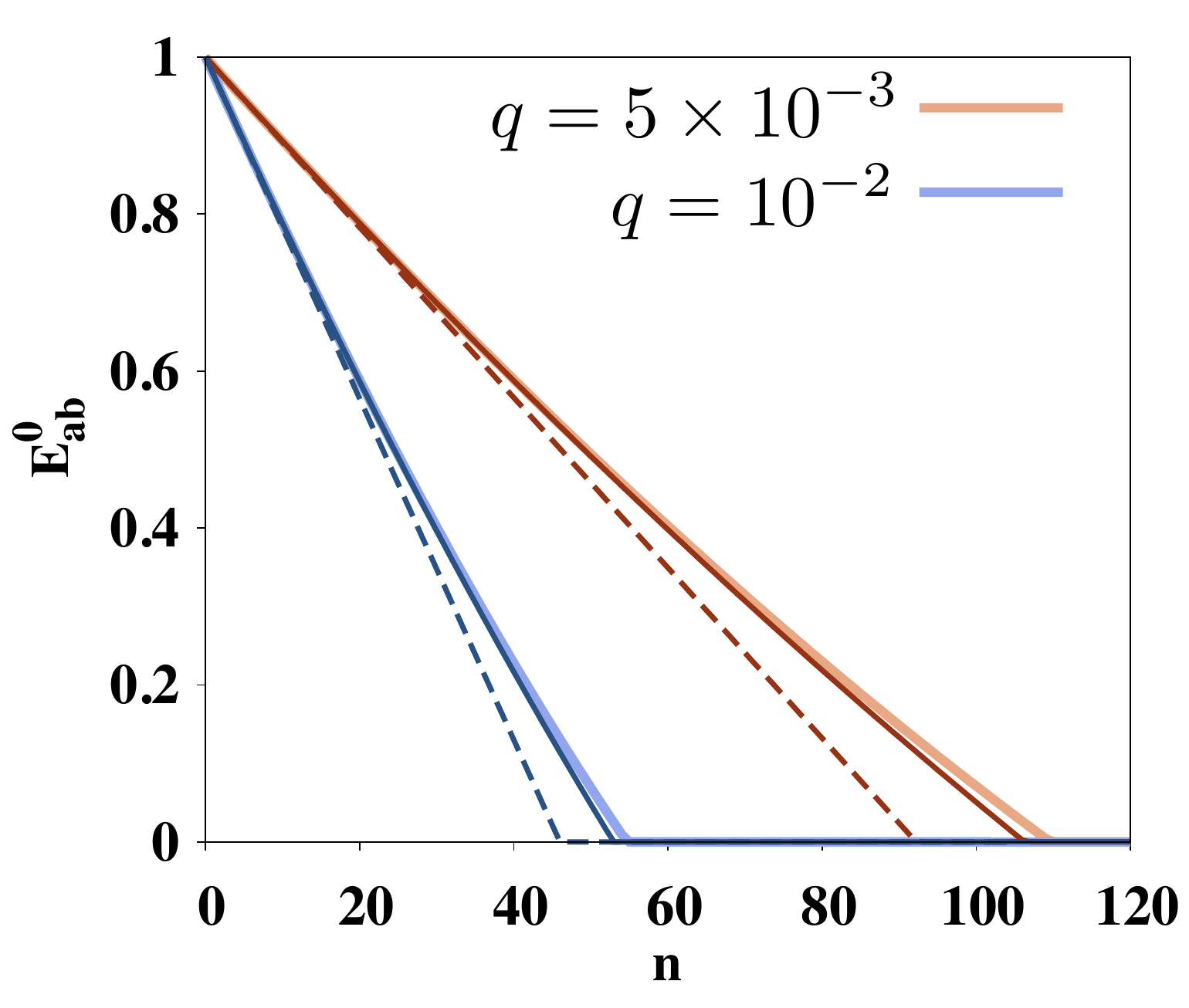}
\caption{(Colour online) \textbf{Measurement-based lower bound as a function of system size.} The variations of $E^0_{ab}$ (Eq.~(\ref{eq:lneg_analytic})) as functions of $n$ for different small values of $q$, with $n_a=n_{ab}=n_{b}=n$. The broken (continuous) lines correspond to the variations of $E^{0}_{ab}$ with $n$ when $E^0_{ab}=\mathcal{O}_0(n)+\mathcal{O}_1(n)$ ($E^0_{ab}=\mathcal{O}_0(n)+\mathcal{O}_1(n)+\mathcal{O}_2(n)$) (see Eq.~(\ref{eq:approx})).}
\label{fig:en}
\end{figure}

In the regime of low noise strengths, $q\rightarrow 0$, upon expanding the logarithm and keeping terms up to second order in $q$, Eq. ~(\ref{eq:lneg_analytic}) leads to  
\begin{eqnarray}
E_{ab}^0&\approx & 1-\frac{3nq}{2\ln 2}+\frac{3n(n-2)q^2}{8\ln 2},\nonumber \\
&=&\mathcal{O}_0(n)+\mathcal{O}_1(n)+\mathcal{O}_2(n),
\label{eq:approx}
\end{eqnarray}
$\mathcal{O}_k(n)$ being the term involving $n$ in order $k$. The variation of $E^0_{ab}$ as a function of $n$ for fixed values of $q$ is depicted in Fig.~\ref{fig:en}, when the noise strength is small. To determine the leading order of $n$ that describes $E^0_{ab}$ for small values of $q$, we plot, in Fig.~\ref{fig:en}, $E^0_{ab}\approx \mathcal{O}_0(n)+\mathcal{O}_1(n)$ (up to first order in $n$, shown by broken line) and $E^0_{ab}\approx \mathcal{O}_0(n)+\mathcal{O}_1(n)+\mathcal{O}_2(n)$ (up to second order in $n$, shown by continuous line) as functions of $n$. It is clear from Fig.~\ref{fig:en} that for a fixed small value of $q$, $E^0_{ab}\approx \mathcal{O}_0(n)+\mathcal{O}_1(n)$ matches the actual variation of $E^0_{ab}$ satisfactorily when $n$ is very small $(\sim 10)$. When $n$ increases, the second order term in $n$ starts to become prominent, and $E^0_{ab}\approx \mathcal{O}_0(n)+\mathcal{O}_1(n)+\mathcal{O}_2(n)$ describes entanglement satisfactorily.

We would like to point out here that the prescription for computing the post-measurement density matrix to obtain a form equivalent to Eq. (\ref{eq:final_form}) remains unchanged for a region $\Omega$ having size larger than two qubits also. The major step in this calculation is the determination of the mixing probabilities according to the general structure of the neighborhood of  $\Omega$ in a graph where $\Omega$ is connected, which can be achieved following procedure similar to that described in this Section and the Appendix \ref{ap:pmstate}.  As mentioned earlier in Sec. \ref{subsec:lb_gdstate}, the main difficulty of estimating localizable multipartite entanglement over a region larger than two qubits in the presence of noise is the lack of computable measures of genuine multipartite entanglement for mixed states. In this paper, we have considered a computable bipartite measure of entanglement, namely, logarithmic negativity, which is equivalent to the genuine multiparty entanglement when $\Omega$ is constituted of two qubits only. However, given a computable multiparty entanglement measure for mixed states, the MLB to the localizable multipartite entanglement over a chosen region $\Omega$ constituted of any number of qubits can, in principle, be computed by following a procedure same as in the case of a two-qubit region. 

 \begin{figure}
\includegraphics[scale=0.45]{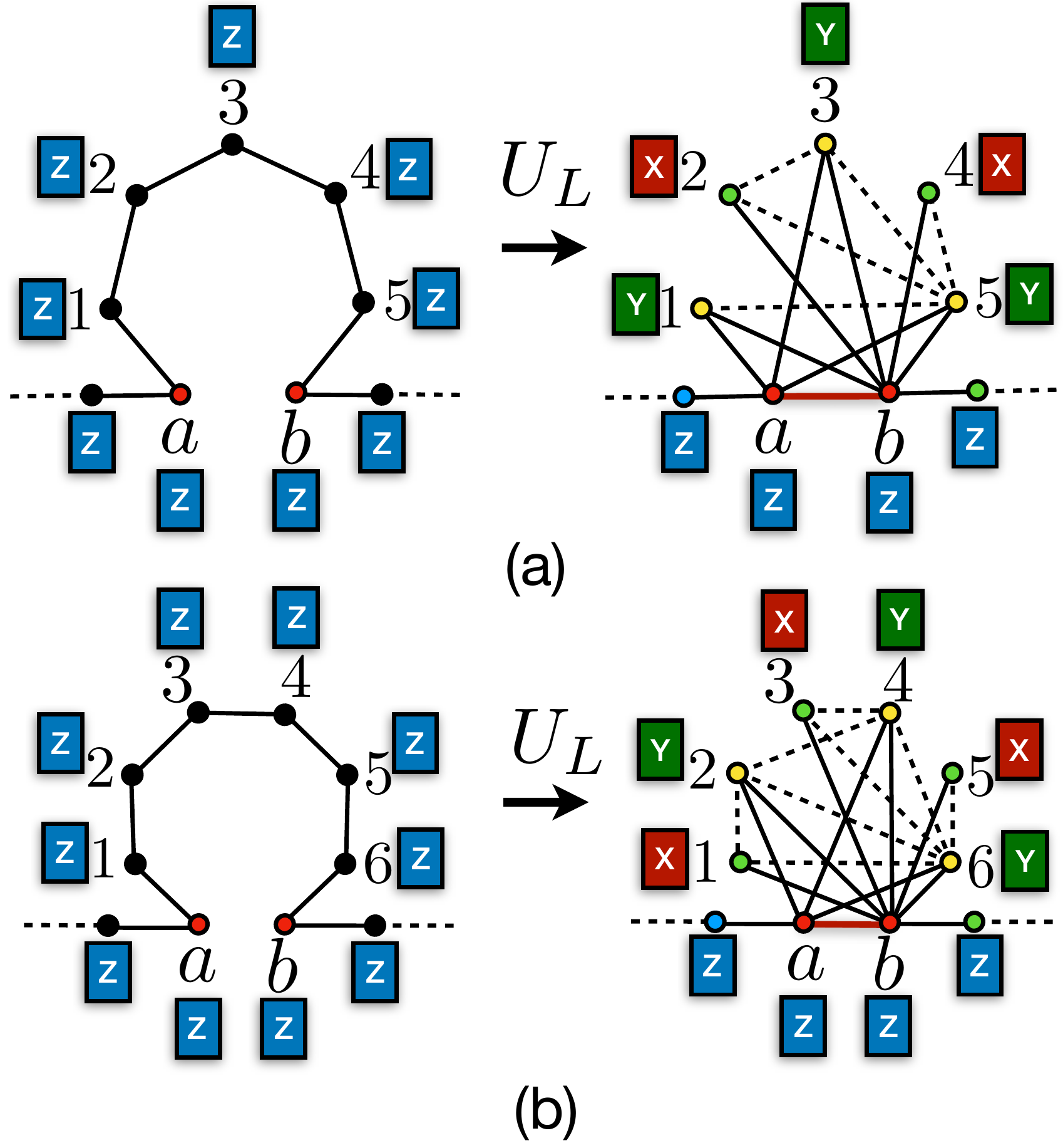}
\caption{(Colour online) \textbf{Schematic representation of local complementation operations on a linear graph under phase-flip noise.} (a) On the left, a linear graph $\mathcal{G}_L$ with two bulk qubits $a$ and $b$, separated by $n_{\mathcal{L}}=5$ qubits, is shown. The noise on each qubit is of PF type ($Z$-type), and is indicated by the labels. A series of LC operations on the qubits in $\mathcal{L}$, given by Eq.~(\ref{eq:lc_linear}), takes $\mathcal{G}_L$ to $\mathcal{G}^\prime$ (on the right), where the link $(a,b)$ exists. The operation also changes the noise on individual qubits according to Eqs.~(\ref{eq:transform})-(\ref{eq:evolve2}) and Appendix \ref{ap:pauli}, which is indicated by the different labels, where label $X$ and $Y$ indicate BF and BPF noise, respectively. (b) A similar transformation is described for a linear graph with $n_{\mathcal{L}}=6$.}
\label{fig:linear_graph}
\end{figure}

\subsubsection*{Linear graph}

We conclude the discussion on the MLB with the example of a linear graph $\mathcal{G}_L$, in which we intend to determine the MLB over two qubits $a$ and $b$, where the total number of qubits along the path connecting $a$ and $b$ is $n_{\mathcal{L}}$. Note here that the qubit pair $(a,b)$ can either be (i) the boundary qubits, so that in $\mathcal{G}_L$, both $\mathcal{N}_a$ and $\mathcal{N}_b$ have size $1$, or they can be (ii) bulk qubits (as in Fig.~\ref{fig:linear_graph}(a)-(b)), where both $\mathcal{N}_a$ and $\mathcal{N}_b$ have size $2$. For the purpose of demonstration, we consider the scenario where $a$ and $b$ are bulk qubits, $n_{\mathcal{L}}\geq 3$, and PF noise is applied to each of the qubits in $\mathcal{G}_L$. The transformation $\mathcal{G}_L\rightarrow\mathcal{G}^\prime$, where $\{a,b\}$ are connected in $\mathcal{G}^\prime$, is constituted of successive LC operations on the qubits in $\mathcal{L}$, starting from the qubit nearest to $a$ and ending at the qubit nearest to $b$ without skipping any qubit in the middle, so that 
\begin{eqnarray}
\mathcal{G}^\prime=\tau_{n_{\mathcal{L}}}\circ\tau_{n_\mathcal{L}-1}\circ\cdots\circ\tau_2\circ\tau_1(\mathcal{G}_L).
\label{eq:lc_linear}
\end{eqnarray} 
The structure of $\mathcal{G}^\prime$ is shown for $n_{\mathcal{L}}=5$ ($n_\mathcal{L}=6$) in Fig.~\ref{fig:linear_graph}(a) (\ref{fig:linear_graph}(b)). The Eq.~(\ref{eq:lc_linear}) can equivalently be represented as $\ket{\mathcal{G}^\prime}=U_L\ket{\mathcal{G}_L}$, with 
\begin{eqnarray}
U_L=U_a\otimes\big(\bigotimes_{i\in\mathcal{L}}V_i\big)\otimes U_b, 
\label{eq:lcu_linear}
\end{eqnarray}
where 
\begin{eqnarray}
U_a&=&(u^z_a)^{n_{\mathcal{L}}},\, U_b=u^z_b,\, V_1=(u^z_1)^{n_{\mathcal{L}}-1}u^x_1,\, V_{n_{\mathcal{L}}}=u^{x}_{n_{\mathcal{L}}}u^z_{n_{\mathcal{L}}},\nonumber \\
V_j&=&(u^z_j)^{n_{\mathcal{L}}-j}u^x_ju^z_j; \,\,2\leq j\leq (n_{\mathcal{L}}-1), 
\label{eq:lcu_linear_def}
\end{eqnarray}
with $u_i^x$ and $u_i^z$ defined in Sec.~\ref{subsec:stab}. Note here that in the case of $n_{\mathcal{L}}=1$, $U_a=u_a^z$, $U_b=u_b^z$, $V_1=u_1^x$, while for $n_{\mathcal{L}}=2$, $U_a=(u_a^z)^2$, $U_b=u_b^z$, $V_1=u_1^zu_1^x$, and $V_2=u_2^xu_2^z$. The transformation of the Pauli operators due to the unitary operators $\{U_a,U_b,V_j;\,j=1,\cdots,n_{\mathcal{L}}\}$ are given in Appendix \ref{ap:pauli}, which describes the change of the type of noise on individual qubits according to Eqs.~(\ref{eq:transform}) and (\ref{eq:evolve2}). The post-LC operation structures of the graphs, as demonstrated in the case of $n_{\mathcal{L}}=5,6,$ in Fig.~\ref{fig:linear_graph}, is such that for $n_{\mathcal{L}}$ odd, $n_a=0$, $n_{ab}=(n_{\mathcal{L}}+1)/2$, and $n_b=(n_{\mathcal{L}}-1)/2$, while for $n_{\mathcal{L}}$ even, $n_a=0$ and $n_{ab}=n_b=n_{\mathcal{L}}/2$. Therefore, $E^0_{ab}$ as a function of $q$ and $n_{\mathcal{L}}$ can be computed by following the methodology discussed in Sec.~\ref{subsec:mlb_arb}. Note here that the values of $n_a$, $n_b$, and $n_{ab}$ in terms of $n_{\mathcal{L}}$ depend on the structure of the graph $\mathcal{G}^\prime$ as well as the noise on the qubits in $\mathcal{N}_{ab}$ in $\mathcal{G}^\prime$. For instance, in the case of the BF noise on all the qubits, irrespective of the value of $n_{\mathcal{L}}$, $n_a=n_{ab}=n_b=1$. The invariance of $E^0_{ab}$ with $n_{\mathcal{L}}$ in the case of BF noise on all the qubits in $\mathcal{G}_{L}$ can be understood by noticing the fact that the optimal measurement basis in the absence of noise corresponds to $X$ measurements on qubits in $\mathcal{L}$, and $Z$ measurements on the rest of the qubits except $a$ and $b$, and the measurement on qubits in $\mathcal{L}$ commutes with the noise.


\section{Conclusions and Outlook}
\label{sec:conclude}

In this paper, we have considered two different approaches of determining computable lower bounds of localizable entanglement for large stabilizer states under noise. One of the approaches is based on local witnesses, whose expectation values can be used to obtain a lower bound of the localizable entanglement. The other approach restricts the allowed directions of the local projection measurements over the qubits outside the specific region of interest over which the localizable entanglement is to be computed. By establishing a relation between the disentangling operation that reduces the full quantum state to the quantum state corresponding to the specific regime, and local $Z$ measurements over qubits outside the region, we have been able to connect these two seemingly different approaches, and have proposed a hierarchy of lower bounds of localizable entanglement. 

Using graph states for demonstration, we show that in the case of graph states exposed to noise, the measurement-based lower bound is greater or equal to the witness-based lower bound. The equality occurs in the case of graph diagonal states, when localizable entanglement over a region constituted of two qubits is to be determined. We have demonstrated how the hierarchy of lower bounds of localizable entanglement is modified due to local unitary transformation, and discussed the behaviour of the lower bounds under physical noise models, such as the local uncorrelated Pauli noise. We have demonstrated that for two-qubit regions, in the case of graph states under local Pauli noise, which form a subset of the complete set of graph-diagonal states, the witness-based lower bound coincides with the measurement-based lower bound. But in the case of three-qubit regions, the measurement-based lower bound is a tighter lower bound for localizable entanglement. We have also proposed an analytical approach to determine the measurement-based lower bound for quantum states of arbitrary size under Pauli noise, and discussed the behaviour of the measurement-based lower bound by performing $Z$-measurement over the qubits outside a two-qubit region as a function of noise strength and system size. The results discussed in this paper are either valid for, or can be translated to more general stabilizer states due to their connection with graph states by local unitary operation. The witness-based lower bounds of localizable entanglement proposed in this paper can be evaluated experimentally without performing a full state tomography, and by considering only one local witness-operator expectation value, which makes it a quantity feasible to be computed in experiments. Also, the measurement-based lower bound discussed in this paper does not require a full optimization with all possible local measurement bases over the qubits outside the region, but needs only local measurement in the computational basis, and can be determined by only knowing the structure of the graph and the type of noise applied to the qubits. Therefore, we expect the quantities and methods introduced in this work to be valuable for the investigation of localizable entanglement in experimental medium- and large-scale noisy stabilizer states.

\acknowledgements

We acknowledge support by U.S. A.R.O. through Grant No. W911NF-14-1-010. The research is also based upon work supported by the Office of the Director of National Intelligence (ODNI), Intelligence Advanced Research Projects Activity (IARPA), via the U.S. Army Research Office Grant No. W911NF-16-1-0070. The views and conclusions contained herein are those of the authors and should not be interpreted as necessarily representing the official policies or endorsements, either expressed or implied, of the ODNI, IARPA, or the U.S. Government. The U.S. Government is authorized to reproduce and distribute reprints for Governmental purposes notwithstanding any copyright annotation thereon. Any opinions, findings, and conclusions or recommendations expressed in this material are those of the author(s) and do not necessarily reflect the view of the U.S. Army Research Office.

\appendix

\section{Optimizing the witness-based lower bound}
\label{ap:wlb_opt}

As discussed in Sec. \ref{subsec:wlb_mlb}, we need to determine the minimum value of negativity that is consistent with experimentally determined expectation values $\{\omega\}$ of local witness operators.  In our case, we only focus on the witness operator $\mathcal{W}_{\Omega}^g$, and the optimization problem aims to find the solution of 
\begin{eqnarray}
	N_{g}^{\mbox{\scriptsize min\normalsize}}=&&\inf \|{\left(\rho_{AB}\right)}^{T_{A}}\|_1-1,\nonumber \\
	\mbox{subject to } && \text{Tr}\left(\rho_{AB} \mathcal{W}_{\Omega}^g\right) = \omega,\nonumber\\
	&& \rho_{AB}\geq 0,\nonumber\\
	&& \text{Tr}\left(\rho_{AB}\right)=1,
	\label{eq:opt1}
\end{eqnarray}
where the optimization is done over all possible states $\rho_{AB}$. Here, we have considered a specific bipartition of the region $\Omega$ into the subparts $A$ and $B$, and $E^{\mbox{\scriptsize min\normalsize}}=N_g^{\mbox{\scriptsize min\normalsize}}$ is the quantity to be computed. Using the variational characterization of trace-norm, and following the procedure described in Ref.~\cite{eisert2007}, one arrives at 
\begin{eqnarray}
 N_{g}^{\mbox{\scriptsize min\normalsize}}\geq && E^{\mathcal{W}}_{\Omega}(\omega)=\inf \text{Tr}\left[ D\left(\rho_{AB}\right)^{T_{A}}\right]-1,\nonumber \\
	 \text{subject to } && \text{Tr}[\rho_{AB}\mathcal{W}_{\Omega}^g] = \omega,\nonumber\\
	&& \rho_{AB}\geq 0,\nonumber\\
	&& \text{Tr}[\rho_{AB}]=1,
	\label{eq:opt2}
\end{eqnarray}
where $D$ is any operator such that $\|D\|_{\infty}=1$, and the right-hand-side of the inequality in Eq.~(\ref{eq:opt2}) provides $E_{\Omega}^{\mathcal{W}}$ corresponding to negativity. Considering $D$ to be of the form $ D=-f\left(\mathcal{W}^{g}_{\Omega}\right) ^{T_{A}}+h I$ involving the partial transpose of the local witness operator that has been measured, where the coefficients $f$ and $h$ are such that $\|D\|_{\infty}=1$, one arrives at a simple form of the lower bound, given by 
\begin{eqnarray}
\hspace{-5mm}E^\mathcal{W}_{\Omega}(\omega)=\underset{f,h}{\max} (-fw+h-1) \text{ subject to } \|D\|_{\infty}=1. 
\label{eq:fh} 
\end{eqnarray}
Note that the form chosen for $D$ allows one to avoid the minimization involved in (\ref{eq:opt2}). Note also that any set of values of $f,h$ subject to $ \|D\|_{\infty}=1 $ provides a value of the lower bound. 

However, we would like to find the best possible value by performing the optimization in Eq. (\ref{eq:fh}). In order to do so, we note that $\left( \mathcal{W}^{g}_{\Omega}\right) ^{T_{A}}=1/2\,I-\rho_{\mathcal{G}_{\Omega}}^{T_{A}}$, and since $\rho_{\mathcal{G}_{\Omega}}^{T_{A}}$ is diagonal in the graph state basis, so is $D$. In the case of a region $\Omega$ of size two, $A$ and $B$ denotes the qubits constituting $\Omega$, and
\begin{eqnarray}
\rho_{\mathcal{G}_{\Omega}}^{T_{A}}&=&\frac{1}{2}\big[Z_0\rho_{\mathcal{G}_{\Omega}}Z_0+Z_1\rho_{\mathcal{G}_{\Omega}}Z_1+Z_2\rho_{\mathcal{G}_{\Omega}}Z_2 -Z_3\rho_{\mathcal{G}_{\Omega}}Z_3\big], \nonumber \\
\end{eqnarray}
following the notation for GD states. In the case of $\Omega$ constituted of three qubits, say, $1$, $2$, and $3$, one can consider three possible bipartitions of $\Omega$, which are equivalent under qubit permutations. For the bipartition $1|23$, one obtains 
\begin{eqnarray}
\rho_{\mathcal{G}_{\Omega}}^{T_{A}}&=&\frac{1}{2}\big[Z_0\rho_{\mathcal{G}_{\Omega}}Z_0+Z_3\rho_{\mathcal{G}_{\Omega}}Z_3+Z_4\rho_{\mathcal{G}_{\Omega}}Z_4 -Z_7\rho_{\mathcal{G}_{\Omega}}Z_7\big]. \nonumber \\
\end{eqnarray}
The singular values of $ D $ are $\{|h|,|h-f|\}$ and $\{|h|,|h-f|,|h-f/2|\}$ for regions of size two and three, respectively. Since $ \left\| D\right\| _{\infty}=1$, the maximum singular value among them must be $1$, which implies $\max\{|h|,|h-f|\}=1$ , because the third singular value is smaller or equal than the first or the second for any pair $ h,f $. This can be satisfied with four sets of solutions of $f$ and $h$, given by (i) ($h=1$, $ 0\leq f \leq 2$), (ii) ($h=-1$,$ -2\leq f \leq 0$), (iii) ($h=1+f$, $ -2\leq f \leq 0  $), and (iv) ($h=-1-f $, $ 0\leq f \leq 2 $). As mentioned earlier, although any of the four pairs of values of $f$ and $h$ provides a valid lower bound for $N_{g}^{\text{min}}$, we choose the best of them. In the case when $\omega<0$, the optimal pair is ($h=1$, $f=2$), from (i), and for $\omega\geq 0$, the optimal values are ($ h=1 $ and $ f=0 $) from (i) and (iii), which leads to 
\begin{eqnarray}
E^\mathcal{W}_{\Omega}(\omega) &=&
      \begin{cases}
      -2\omega, & \text{for } \omega < 0,  \\
      0, & \text{for } \omega \geq 0,
      \end{cases}. 
\end{eqnarray}
The lower-bound corresponding to the logarithmic negativity also can now be straightforwardly obtained from the value of $E^\mathcal{W}_{\Omega}(\omega)$ by using Eq.~(\ref{eq:logneg}).

\section{Determination of the mixing probabilities}
\label{ap:pmstate}

Here we present the crucial steps of the derivation of the forms of $q_{\overline{\Omega}}^\beta$, given in Eq.~(\ref{eq:mix_prob}). For the purpose of demonstration, let us consider the correction $\mathcal{Z}^{0}_{ab}=I_a\otimes I_b$. Let us assume that the number of ``$1$"s in the outcome $k^\prime\equiv k_{r_1}k_{r_2}k_{r_3}\cdots k_{r_{N-2}}$, where $r_i\in\tilde{\mathcal{N}}_a^1$ is $m_a^1$, and we use similar notations for the sets $\tilde{\mathcal{N}}_b^1$ and $\tilde{\mathcal{N}}_{ab}^1$. According to Eqs.~(\ref{eq:meas_rule_gen}) and (\ref{eq:form_projectors_2_single}), the correction $\mathcal{Z}^{0}_{ab}$ may result iff (i) $m_a^1$, $m_{ab}^1$, and $m_b^1$ are all odd, or (ii) all even. The value of $k_{r_i}^\prime=1$ for $r_i\in\tilde{\mathcal{N}}_a^1$ when (a) $k_{r_i}=0$ is changed to $k_{r_i}^\prime=1$, due to the application of a noise of \textbf{Type 1} with probability $s$ ($0\leq s\leq 1$), and when (b) $k_{r_i}=1$ remains unchanged with a probability $(1-s)$. Let us denote the number of occurrences of event (a) by $m_a^{01}$, and the same for event (b) by $m_{a}^{11}$, where $m_a^{01}+m_a^{11}=m_a^1$. Similar descriptions can also be adopted for qubits in $\tilde{\mathcal{N}}_{b}^1$ and $\tilde{\mathcal{N}}_{ab}^1$.  An odd value of $m_a^1$ may result either when (1) $m_a^{01}$ is odd and $m_a^{11}$ is even, or when (2) $m_a^{01}$ is even and $m_a^{11}$ is odd. The probability of occurrence of the event (1) is $P_{(1)}=P_{(1)}^1P_{(1)}^2$, where 
\begin{eqnarray}
P_{(1)}^1&=&\sum_{m_a^{01}=1,3,5,\cdots}\binom{n_a^0}{m_a^{01}} s^{m_a^{01}}(1-s)^{n_a^0-m_a^{01}},\nonumber \\
P_{(1)}^2&=&\sum_{m_a^{11}=0,2,4,\cdots}\binom{n_a^1}{m_a^{11}} (1-s)^{m_a^{11}}s^{(n_a^1-m_a^{11})}.
\end{eqnarray}
Similarly, for the event (2), $P_{(2)}=P_{(2)}^1P_{(2)}^2$, where 
\begin{eqnarray}
P_{(2)}^1&=&\sum_{m_a^{01}=0,2,4,\cdots}\binom{n_a^0}{m_a^{01}} s^{m_a^{01}}(1-s)^{n_a^0-m_a^{01}},\nonumber \\
P_{(2)}^2&=&\sum_{m_a^{11}=1,3,5,\cdots}\binom{n_a^1}{m_a^{11}} (1-s)^{m_a^{11}}s^{(n_a^1-m_a^{11})}.
\end{eqnarray}
These expressions can be simplified by using the following identities, where $0\leq t\leq 1$.
\begin{eqnarray}
\frac{1}{2}[1+(1-2t)^N]&=&\sum_{m=0,2,4,\cdots}\binom{N}{m}t^m(1-t)^{N-m},\nonumber \\
\frac{1}{2}[1-(1-2t)^N]&=&\sum_{m=1,3,5,\cdots}\binom{N}{m}t^m(1-t)^{N-m}. 
\label{eq:id}
\end{eqnarray}
Using these identities, the probability that $m_a^1$ is odd is obtained as 
\begin{eqnarray}
P_a^-&=&P_{(1)}+P_{(2)}\nonumber \\
&=&\frac{1}{2}[1-(-1)^{n_a^1}(1-2s)^{n_a}].
\end{eqnarray}
A similar approach for the probability of obtaining an even value of $m_a^1$ leads to 
\begin{eqnarray}
P_a^+&=&\frac{1}{2}[1+(-1)^{n_a^1}(1-2s)^{n_a}].
\end{eqnarray}
In analogy, the corresponding probabilities in the case of $\tilde{\mathcal{N}}_b^1$ and $\tilde{\mathcal{N}}_{ab}^1$ are obtained as 
\begin{eqnarray}
P_b^\pm &=&\frac{1}{2}[1\pm (-1)^{n_b^1}(1-2s)^{n_b}],\nonumber \\
P_{ab}^\pm &=&\frac{1}{2}[1\pm (-1)^{n_{ab}^1}(1-2s)^{n_{ab}}].
\end{eqnarray}
Therefore, the probability with which a correction $\mathcal{Z}^0_{ab}$ is applied on the state $\rho_{\mathcal{G}_{ab}}$ can be written as 
\begin{eqnarray}
q_{\overline{\Omega}}^{0} = P_{a}^- P_{ab}^-P_{b}^- + P_{a}^+ P_{ab}^+ P_{b}^+,
\end{eqnarray}
which provides the mixing probability corresponding to the state $\Lambda_{ab}(\rho_{\mathcal{G}_{ab}}^0)$ in the state ${\rho^\prime}^k_{ab}$. Similarly, the expressions for $q_{\overline{\Omega}}^\beta$, $\beta=1,2,3$, corresponding to the corrections $\mathcal{Z}_{ab}^1=I_a\otimes Z_b$,  $\mathcal{Z}_{ab}^2=Z_a\otimes I_b$, and  $\mathcal{Z}_{ab}^3=Z_a\otimes Z_b$, can also be obtained as 
\begin{eqnarray}
q_{\overline{\Omega}}^{1} &=& P_{a}^- P_{ab}^-P_{b}^+ + P_{a}^+ P_{ab}^+ P_{b}^-,\nonumber\\
q_{\overline{\Omega}}^{2} &=& P_{a}^+ P_{ab}^-P_{b}^- + P_{a}^- P_{ab}^+ P_{b}^+,\nonumber\\
q_{\overline{\Omega}}^{3} &=& P_{a}^- P_{ab}^+P_{b}^- + P_{a}^+ P_{ab}^- P_{b}^+.
\end{eqnarray}
We point out here that according to the convention used in the paper (Eq.~(\ref{eq:channel_prob})), the probability $s=q/2$ in the case of the BF, the BPF, and the DP channels, while in the case of the PF channel, $s=0$.

\section{Transformation of Pauli operators in linear graph}
\label{ap:pauli}
The transformation $\ket{\mathcal{G}^\prime}=U_L\ket{\mathcal{G}_L}$ of the graph state $\ket{\mathcal{G}}$, corresponding to the transformation of linear graph given in Eq.~(\ref{eq:lc_linear}), is determined by the local unitary operator 
$U_L=U_a\otimes\big(\bigotimes_{i\in\mathcal{L}}V_i\big)\otimes U_b$, where $U_a=(u^z_a)^{n_{\mathcal{L}}}$, $U_b=u^z_b$, $V_1=(u^z_1)^{n_{\mathcal{L}}-1}u^x_1$, $V_{n_{\mathcal{L}}}=u^{x}_{n_{\mathcal{L}}}u^z_{n_{\mathcal{L}}}$, and $V_j=(u^z_j)^{n_{\mathcal{L}}-j}u^x_ju^z_j$ for $2\leq j\leq (n_{\mathcal{L}}-1)$,  with $u_j^x=\exp[(-\text{i}\pi/4)X_j]$ and $u_j^z=\exp[(\text{i}\pi/4)Z_j]$. The transformation of the Pauli operators due to the unitary operators $\{U_a,U_b,V_j;\,j=1,\cdots,n_{\mathcal{L}}\}$ are given by 
\begin{eqnarray}
U_a X_a U_a^{-1}&=& 
      \begin{cases}
      (-1)^mX_a, & \text{for } n_{\mathcal{L}}=2m  \\
      (-1)^{m+1}Y_a, & \text{for } n_{\mathcal{L}}=2m+1
      \end{cases},\nonumber \\
U_a Y_a U_a^{-1}&=& 
      \begin{cases}
      (-1)^mY_a, & \text{for } n_{\mathcal{L}}=2m  \\
      (-1)^{m}X_a, & \text{for } n_{\mathcal{L}}=2m+1
      \end{cases},\nonumber \\
U_a Z_a U_a^{-1}&=& Z_a,
\end{eqnarray}
\begin{eqnarray}       
U_b X_b U_b^{-1}&=& -Y_b,\nonumber \\
U_b Y_b U_b^{-1}&=& X_b,\nonumber \\
U_b Z_b U_b^{-1}&=& Z_b,
\end{eqnarray} 
\begin{eqnarray}
V_1 X_1 V_1^{-1}&=& 
      \begin{cases}
      (-1)^mY_1, & \text{for } n_{\mathcal{L}}=2m  \\
      (-1)^{m}X_1, & \text{for } n_{\mathcal{L}}=2m+1
      \end{cases},\nonumber \\
V_1 Y_1 V_1^{-1}&=& Z_1,\nonumber \\
V_1 Z_1 V_1^{-1}&=& 
      \begin{cases}
      (-1)^mX_1, & \text{for } n_{\mathcal{L}}=2m  \\
      (-1)^{m+1}Y_1, & \text{for } n_{\mathcal{L}}=2m+1
      \end{cases},
\end{eqnarray}
\begin{eqnarray}       
V_{n_{\mathcal{L}}} X_{n_{\mathcal{L}}} V_{n_{\mathcal{L}}}^{-1}&=& -Z_{n_{\mathcal{L}}},\nonumber \\
V_{n_{\mathcal{L}}}  Y_{n_\mathcal{L}} V_{n_{\mathcal{L}}} ^{-1}&=& X_{n_{\mathcal{L}}},\nonumber \\
V_{n_{\mathcal{L}}} Z_{n_{\mathcal{L}}} V_{n_{\mathcal{L}}} ^{-1}&=& -Y_{n_{\mathcal{L}}},
\end{eqnarray} 
and 
\begin{eqnarray}
V_j X_j V_j^{-1}&=& -Z_j,\nonumber \\
V_j Y_j V_j^{-1}&=& 
      \begin{cases}
      (-1)^mX_j, & \text{for } n_{\mathcal{L}}-j=2m  \\
      (-1)^{m+1}Y_j, & \text{for } n_{\mathcal{L}}-j=2m+1
      \end{cases},\nonumber \\
V_j Z_j V_j^{-1}&=& 
      \begin{cases}
      (-1)^{m+1}Y_j, & \text{for } n_{\mathcal{L}}-j=2m  \\
      (-1)^{m+1}X_j, & \text{for } n_{\mathcal{L}}-j=2m+1
      \end{cases}, 
\end{eqnarray}
where $m=0,1,2,\cdots$ if $n_{\mathcal{L}}$ or $n_{\mathcal{L}}-j=2m+1$, and  $m=1,2,3,\cdots$ if $n_{\mathcal{L}}$ or $n_{\mathcal{L}}-j=2m$.


\end{document}